\documentclass[sort&compress]{elsarticle}
\pdfoutput=1

\usepackage{hyperref}
\usepackage{algorithm}
\usepackage[noend]{algorithmic} 

\usepackage{subfigure}
\usepackage{amssymb}
\usepackage{ulem}

\newcommand{\daniel}[1]{}
\newcommand{\kamel}[1]{}
\newcommand{\owen}[1]{}

\newcommand{\cutmarch}[1]{}

\newcommand{\notessential}[1]{#1}
\newcommand{\altnotessential}[1]{#1}   

\renewcommand{\notessential}[1]{\textcolor{red}{[candidate for reentry:]}#1\textcolor{red}{[/end]}}

\newcommand{\omitproof}[1]{#1}

\newcommand{\cut}[1]{{[\textcolor{green}{was cut}]{#1}}[\textcolor{green}{/end}]}
\newcommand{\wascut}[1]{{[\textcolor{green}{was cut}]{#1}}[\textcolor{green}{/end}]}
\newcommand{\killforgood}[1]{}
\newcommand{\suggestwecut}[1]{}
\usepackage{verbatim}
\newcommand{\lxor}{\oplus}
\newtheorem{theorem}{Theorem}  
\newtheorem{proposition}{Proposition}  
\newtheorem{corollary}{Corollary}  
\newtheorem{lemma}{Lemma}  
\newtheorem{definition}{Definition}  
\newproof{pf}{Proof}
\def\qed{\relax\ifmmode\hskip2em \Box\else\unskip\nobreak\hskip1em $\Box$\fi}
\usepackage{graphicx}

\graphicspath{{.}{../BitmapIndexCpp/dbash/dolap2008experiments/}{../BitmapIndexCpp/dbash/dolap2008experiments/syntheticcolumnreordering/}{../BitmapIndexCpp/dbash/dolap2008experiments/syntheticqueries/}{../BitmapIndexCpp/synthetic/}{../BitmapIndexCpp/bash/dolap08/kingjames4d/}{../BitmapIndexCpp/bash/block_usincome/}{../BitmapIndexCpp/bash/block_dbgen/}{../BitmapIndexCpp/bash/block_netflix/}{../BitmapIndexCpp/bash/}{../BitmapIndexCpp/scripts/}}

\usepackage{ifpdf}
\usepackage{color}
\ifpdf
\else
\fi
\usepackage{url}
\journal{Data \& Knowledge Engineering}

\begin{document}
\begin{frontmatter}


\title{
Sorting improves word-aligned bitmap indexes}

 \author[UQAM]{Daniel Lemire\corref{cor1}} \ead{lemire@acm.org} 
 \author[UNB]{Owen Kaser} \ead{o.kaser@computer.org}  
\author[UQAM]{Kamel Aouiche}\ead{kamel.aouiche@gmail.com}

 \address[UQAM]{\scriptsize LICEF, Universit\'e du Qu\'ebec \`a Montr\'eal (UQAM), 100 Sherbrooke West, Montreal, QC, H2X 3P2 Canada
}

 \address[UNB]{\scriptsize Dept.\ of CSAS, University of New Brunswick, 100 Tucker Park Road, Saint John, NB, Canada}
 \cortext[cor1]{Corresponding author. Tel.: 00+1+514 987-3000 ext. 2835; fax: 00+1+514 843-2160.} 

\begin{abstract}
Bitmap indexes must be compressed to reduce input/output costs and minimize CPU usage.
To accelerate logical operations (AND, OR, XOR) over bitmaps, we use techniques
based  on run-length encoding (RLE), such
as Word-Aligned Hybrid (WAH) compression.
These techniques are sensitive to the order of the rows: a simple lexicographical
sort can divide the index size by 9 and make indexes several times faster.
We investigate row-reordering heuristics. 
Simply permuting the columns of the table
  can increase 
 the  sorting efficiency by 40\%.
 Secondary contributions include efficient algorithms to construct and aggregate bitmaps.
 The effect of word length is also reviewed by constructing 16-bit, 32-bit and 64-bit indexes.  Using 64-bit CPUs, we find that 64-bit indexes are
 slightly faster than 32-bit indexes despite being nearly twice as large. 
\end{abstract}

\begin{keyword}
Multidimensional Databases \sep Indexing  \sep Compression \sep  Gray codes
\end{keyword}
\end{frontmatter}

\section{Introduction}

Bitmap indexes are among the most commonly used indexes in data warehouses~\cite{bellatreche2007sap,davis2007idw}.
Without compression, bitmap indexes can be impractically large and slow.
Word-Aligned Hybrid (WAH)~\cite{wu2006obi} is a competitive compression technique:
compared to LZ77~\cite{chan1998bid} and Byte-Aligned Bitmap Compression (BBC)~\cite{874730},
WAH indexes can be ten times faster~\cite{502689}.

Run-length encoding (RLE) and similar encoding schemes (BBC and WAH) make it possible to compute
logical operations between bit\-maps in
 time proportional to the compressed size of
the bitmaps.
However, their efficiency depends on the order of the rows.  
While we show 
that
computing the best order\owen{prefer to : ordering} is NP-hard,
simple heuristics such as lexicographical sort are effective.

\begin{table}
\caption{\label{table:marketing}Comparison between the current paper and related work}
\begin{tabular}{p{3cm}cp{3.3cm}p{2cm}}
\hline
reference & largest index  & reordering heuristics & metrics \\
 &  (uncompressed) &  &  \\ \hline
Sharma \& Goyal~\cite{sharma2008emc} & $6\times 10^7$~bits  & Gray-code & index size \\[0.5ex]
Apaydin et al.~\cite{apay:data-reordering} & --- na ---      & Lexicographical, Gray-code & runs\\[3.5ex]
Pinar et al.~\cite{pinar05}, Canahuate et al\cite{pinarunpublished}&  $2\times 10^9$~bits  & Gray-code, na\"\i{}ve 2-switch, bitmaps sorted by set bits or compressibility & index size, query speed \\[6.5ex]
current paper & $5\times 10^{13}$~bits & Lexicographical,  Gray-code, Gray-Frequency, Frequent-Component,  partial (block-wise) sort,  column and bitmap reorderings &
{index size,  construction time, query speed} \\ \hline
\end{tabular}
\end{table}

Table~\ref{table:marketing} compares the current paper to related work.
Pinar et al.~\cite{pinar05}, Sharma and Goyal~\cite{sharma2008emc}, and Canahuate et al.~\cite{pinarunpublished} used \suggestwecut{Gray-code}
row sorting to improve RLE and WAH compression.
However, their largest bitmap index 
 could
fit uncompressed in RAM on a PC\@. Our data sets are 1~million times larger.

Our main contribution is an evaluation of
\killforgood{practical histogram-oblivious and histogram-aware} heu\-ris\-tics for the row ordering problem over large data sets.
Except for the na\"\i{}ve 2-switch heuristic, we review all previously known heuristics, and we consider several novel
heuristics including  lexicographical ordering, Gray-Frequency, partial sorting, and column   reorderings.
Because we consider large data sets, we can meaningfully address the index construction time.
Secondary contributions include 
 \begin{itemize}
\item  guidelines about when
``unary''
bitmap  encoding is preferred (\S~\ref{sec:Guidelinesfork});
 \item an improvement over
the na\"{\i}ve bitmap construction algorithm---it is now practical to construct
bitmap indexes over tables with hundreds of millions of rows and millions of attribute values (see Algorithm~\ref{algo:owengenbitmap});
 \item an algorithm to compute important Boolean operations over many bitmaps in time $ O( (\sum_{i=1}^L \vert B_i \vert)  \log L)$ where $\sum_{i=1}^L \vert B_i \vert$ is the total size of the bitmaps (see Algorithm~\ref{algo:genrunlengthmultiplefaster});
 \item the observation that 64-bit indexes can be slightly faster than 32-bit indexes on a 64-bit CPU, despite file sizes nearly twice as large (see \S~\ref{sec:rangequeries-exper}).
 \end{itemize}
The last two contributions are extensions of the conference version of this paper~\cite{kaserdolap2008}. 

The remainder of this paper is organized as follows.  
We define bitmap indexes in \S~\ref{sec:bitmapIndexes}, where we also
explain how to
map attribute values to bitmaps using encodings such as $k$-of-$N$. 
We present compression techniques in \S~\ref{sec:compression}.
In \S~\ref{sec:FindingthebestreorderingisNPHard},
we 
consider the complexity of the row-reordering problem.  Its NP-hardness
motivates use of fast heuristics, and 
in \S~\ref{sec:SortingtoImproveCompression}, we review
sorting-based heuristics.
In \S~\ref{sec:multi}, we analyze $k$-of-$N$ encodings
further to determine the best possible encoding.
Finally, \S~\ref{sec:Experiment} reports on several experiments.

\section{Bitmap indexes}\label{sec:bitmapIndexes}

We find bitmap indexes in several database systems, apparently beginning
with the MODEL~204 engine, commercialized for the IBM~370 in 
1972~\cite{658338}. 
Whereas it is commonly reported~\cite{hammer2003cea} that bitmap indexes
are suited to small dimensions such as gender or marital status,  they
also work over large dimensions~\cite{oraclevivekbitmap,wu2006obi}. 
And as the number of dimensions increases, bitmap indexes become competitive
against  specialized multidimensional index structures such as R-trees~\cite{671192}.

The simplest and most common method of bitmap indexing associates a bitmap with every
attribute value $v$ of every attribute $a$;
the bitmap represents the predicate $a=v$. Hence, the list
\texttt{cat,dog,cat,cat,bird,bird}
becomes the three bitmaps 1,0,1,1,0,0, 0,1,0,0,0,0,  and 0,0,0,0,1,1.
 For a table with $n$~rows (facts) and $c$~columns (attributes/dimensions), 
each bitmap has length $n$;
initially, all bitmap values are set to 0.
Then, for row $j$, we set the $j^{\mathrm{th}}$~component of $c$~bitmaps to 1.
 If the $i^{\mathrm{th}}$~attribute has $n_i$ possible values, we have
 $L = \sum_{i=1}^{c} n_i$~bitmaps.


We expect the number of bitmaps in \owen{was: a bitmap} an
index to be smaller than the
number of rows. They are equal if we index a row identifier  using
a unary bitmap index. 
However,
we typically find frequent attribute values~\cite{aouiche2007cfp}. 
For instance,
in a Zipfian collection of $n$~items with $N$~distinct values,
the item of rank $k\in\{1,\ldots,N\}$ 
occurs with frequency $\frac{n/k^s}{\sum_{j=1}^{N} 1/j^s}$.
The least frequent item has
frequency $\frac{n/N^s}{\sum_{j=1}^{N} 1/j^s}$ and 
we have that $\sum_{j=1}^{N} 1/j^s \geq 1$. 
Setting   $\frac{n/N^s}{\sum_{j=1}^{N} 1/j^s}\geq 1$
and assuming $N$ large, we have
$N^s  \leq n$,
so that $N \leq \sqrt[s]{n}$. Hence, for highly
skewed distributions ($s\geq 2$), the number of
distinct attribute values $N$ is much smaller than the
number of rows $n$.

Bitmap indexes
are fast, because we find rows having a given value $v$ for attribute $a$ by
reading only the bitmap
corresponding to value $v$ (omitting the other bitmaps for attribute $a$), and there is
only one bit (or less, with compression) to process for each row.
 More complex queries are achieved
with logical operations (AND, OR, XOR, NOT) over bitmaps and
current microprocessors can do 32~or 64~bitwise operations 
in a single machine instruction. 

Bitmap indexes can be highly compressible: 
for row $j$, exactly one bitmap per column will have its  $j^{\mathrm{th}}$ entry set to 1. Although the entire index 
has $nL$~bits, there are only $nc$~1's; for many tables, $L \gg c$ and thus the average table is very sparse.  
Long (hence compressible) 
runs of 0's are expected.

Another approach to achieving small indexes is to
reduce the number of bitmaps for large dimensions. 
Given
$L$~bitmaps, there are $L(L-1)/2$~\emph{pairs} of bitmaps.   So, instead of 
mapping an attribute value to a single bitmap, we map them to pairs of bitmaps  (see Table~\ref{tab:examples1ofk}). 
We refer to this technique 
as 2-of-$N$ encoding~\cite{wong1985btf};
 with it, we can use
far fewer bitmaps for large dimensions.  For instance, with only 2\,000 bitmaps,
we can represent an attribute with 2~million distinct values.
Yet 
the average
bitmap density is much higher with 2-of-$N$ encoding, and thus compression
may be less effective.
More generally, $k$-of-$N$ encoding allows $L$~bitmaps to represent $L \choose k$
distinct values; conversely, using $L=\lceil k n_i^{1/k}\rceil$~bitmaps is sufficient to
 represent $n_i$~distinct values.
 However, searching for a specified value $v$ no longer
involves
scanning a single bitmap.  Instead, the corresponding $k$~bitmaps must
be combined with a bitwise AND\@.  
There is a tradeoff between index size and the
index speed.  

For small dimensions,
using $k$-of-$N$ encoding may fail to reduce the number
of bitmaps, but still reduce the performance.
For example, we have that $N> {N \choose 2}> {N \choose 3}> {N \choose 4}$ for $N\leq 4$,
so that     1-of-$N$ is preferable when $N\leq 4$.
We choose to limit 3-of-$N$ encoding for when $N\geq 6$ and
4-of-$N$ for when $N\geq 8$.
 Hence, we apply
the following heuristic. Any column with less than 5~distinct values is limited
to 1-of-$N$ encoding (simple or unary  bitmap). Any column with less than 21~distinct values,
is limited to $k\in \{ 1,2\}$, and any column with less than 85~distinct values is limited
to $k\in \{1,2,3\}$.

\begin{table}
\caption{\label{tab:examples1ofk}Example of 1-of-N and 2-of-N encoding}
\centering
\begin{tabular}{lcc}
Montreal  & 100000000000000  & 110000 \\
Paris     & 010000000000000  & 101000 \\
Toronto   & 001000000000000  & 100100 \\
New York  & 000100000000000  & 100010 \\ 
Berlin    & 000010000000000  & 100001 
\end{tabular}
\end{table}

Multi-component 
encoding~\cite{chan1998bid} works similarly to $k$-of-$N$ encoding \owen{added, to tie better into concept flow} in
reducing the number of bitmaps: we factor the number of attribute values $n$---or a number slightly exceeding it--- 
as $n=n_1 n_2 \ldots n_\kappa$, with $n_i>1$ for all $i$.
Any number $i \in \{0,1,\ldots, n-1\}$ can be written uniquely in a mixed-radix form as
$i= r_1+ q_1 r_2 + q_1 q_2 r_3+\cdots + r_k q_1 q_2 \ldots q_{\kappa-1}$ where $r_i\in\{0,1,\ldots, q_i - 1\}$. 
 We use a particular encoding scheme (typically 1-of-$N$)
for each of the $\kappa$ values $r_1, r_2,\ldots, r_{\kappa}$
representing the $i^{\mathrm{th}}$~value.
Hence, using $\sum_{i=1}^{\kappa} q_i$~bitmaps we can code $n$~different values.
 Compared
to $k$-of-$N$ encoding, multi-component 
encoding may generate more bitmaps.

\begin{lemma}Given the same number of attribute values $n$, $k$-of-$N$ encoding
 never uses more
 bitmaps than multi-component indexing.
\end{lemma}
\begin{pf}
Consider a $q_1,q_2,\ldots,q_\kappa$-component index. It supports
up to $n= \prod_{i=1}^{\kappa} q_i$ distinct attribute values using
$\sum_{i=1}^{\kappa} q_i$ bitmaps. For $n= \prod_{i=1}^{\kappa} q_i$ fixed, we have that 
$\sum_{i=1}^{\kappa} q_i$ is minimized when $q_i=\sqrt[\kappa]{n}$ for all $i$, hence $ \sum_{i=1}^{\kappa} q_i \geq \lceil \kappa \sqrt[\kappa]{n}\rceil$.
Meanwhile, ${N \choose \kappa} \geq (N/\kappa)^{\kappa}$; hence, by picking
$N=\lceil \kappa \sqrt[\kappa]{n}\rceil$, we have ${N \choose \kappa} \geq n$.
Thus, with at most $\sum_{i=1}^{\kappa} q_i$ bitmaps we can represent at least $n$ distinct values using
$k$-of-$N$ encoding ($k = \kappa$, $N=\lceil \kappa \sqrt[\kappa]{n}\rceil$), which 
shows the result.
\end{pf}

To further reduce the size of bitmap indexes, we can bin the
attribute values~\cite{354819,1155030,stockinger2004esb,1183529}.
For range queries, Sinha and Winslett
use hierarchical binning~\cite{1272746}. 

\section{Compression}\label{sec:compression}

RLE  
compresses \suggestwecut{efficiently
when there are}%
 long runs of
identical values: it replaces any repetition by the number
of repetitions followed by the value being repeated. For example, 
the sequence  11110000 becomes 4140. The counter values (e.g., 4) 
can be stored using  variable-length counters such as gamma~\cite{arxiv:0811.2904} 
or delta codes. With these codes, 
any number $x$ can be written using $O(\log x)$~bits. Alternatively, 
we can used fixed-length counters such as 32-bit integers.
It is common to omit the counter
for single values, and repeat the value twice whenever a counter
is upcoming: e.g., 1011110000 becomes 10114004.

\begin{figure}
\centering
{%
	\includegraphics[width=0.85\columnwidth]{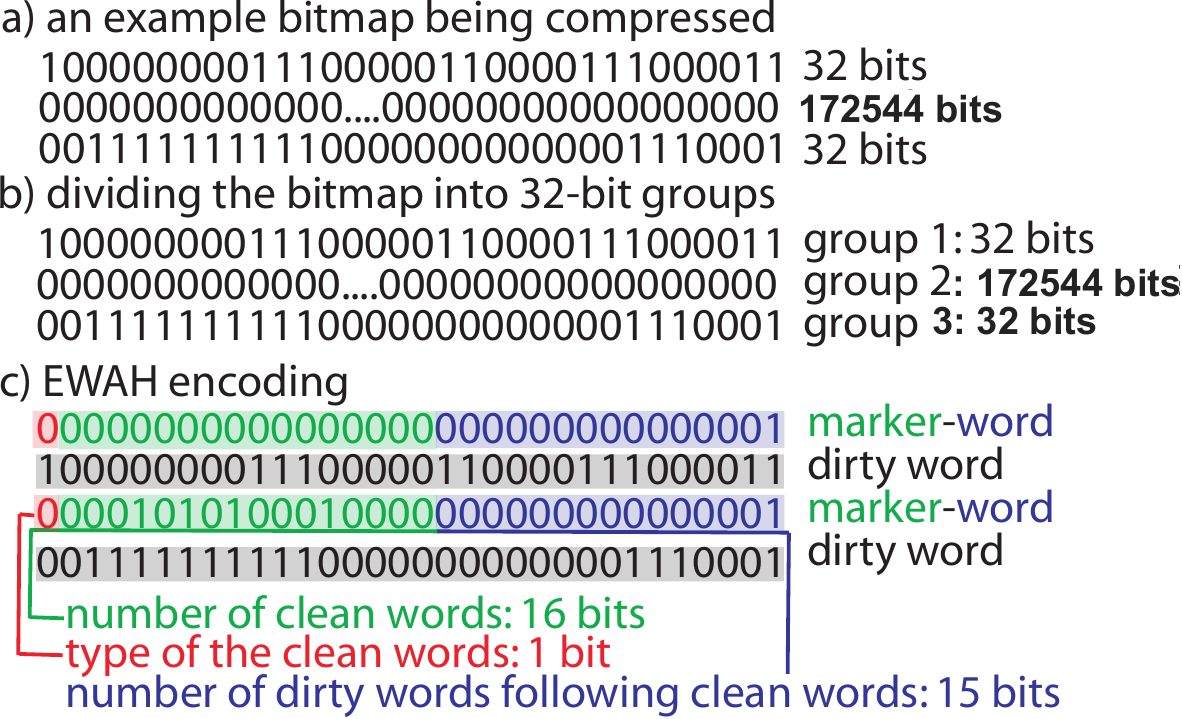}
}
\caption{\label{EWAHfig}Enhanced Word-Aligned Hybrid (EWAH)}
\end{figure}

Current microprocessors
perform operations over words of  32 or 64 bits and not individual bits. Hence,
the CPU cost of RLE might be large~\cite{stockinger2002spa}.
By trading some compression for more speed, Antoshenkov~\cite{874730} defined a RLE variant working over bytes
instead of bits (BBC). 
Trading even more compression for even more speed,
Wu et al.~\cite{wu2006obi} proposed WAH. 
Their scheme is made of two different types of words\footnote{For
simplicity, we limit our exposition to 32~bit words.}. 
The first bit of every word is true (1) for a 
running sequence of 31-bit \emph{clean} words (0x00 or 1x11),
and false (0) for a verbatim (or \emph{dirty}) 31-bit word.
Running sequences are  stored using 1~bit to distinguish between the type
of word (0 for 0x00 and 1 for 1x11) and 30~bits to represent the number of consecutive
clean words. 
Hence, a bitmap of length 62 containing a single 1-bit at position $32$ 
would be coded as
the words 100x01 and 010x00.  
Because dirty words are stored in units of 31~bits using 32~bits, 
WAH compression can expand the data by 3\%.
We studied a WAH variant
that we called Enhanced Word-Aligned Hybrid (EWAH):
in a technical report, Wu et al.~\cite{Wu2001} called the same scheme Word-Aligned Bitmap Code (WBC).
Contrary to WAH compression, EWAH may never (within 0.1\%) generate a compressed
bitmap larger than the uncompressed bitmap.
It also uses only two types of words (see Fig.~\ref{EWAHfig}), where the first type is a 32-bit verbatim word. The second type of word is a
marker word: the first bit  indicates which clean word will follow,
half the bits (16~bits) are used to store the number of clean words, and the rest of the bits (15~bits) are used to store the number of dirty words following the clean words.
EWAH  bitmaps begin with a marker word.
\subsection{Comparing WAH and EWAH}
Because EWAH uses only 16~bits to store the number of clean words, it may be less efficient
than WAH when there are many consecutive sequences of  $2^{16}$~identical clean words. 
The seriousness of this problem is limited because
tables   are indexed
in blocks of rows which fit in RAM: the length of runs does
not grow without bounds even if the table does. 
In \S~\ref{subsection:efficiencyofEWAH}, we show
that this overhead on compressing clean words is at most 14\%
on
our sorted data sets---and this percentage is much lower (3\%) when considering
only unsorted tables. 
Furthermore,
about half of
the compressed bitmaps are made of dirty words, on which EWAH is 3\% more efficient
than WAH. \owen{flipped around to avoid a ``where'' I disliked.}

We can alleviate this compression overhead over clean words in several ways. On the one hand,  we can
allocate 
more than half of the bits to encode the runs of clean words~\cite{Wu2001}. On the other hand, when a marker word indicates a run of $2^{16}$~clean words, we could use
the convention that the next word  indicates the number of remaining clean words. Finally, this compression penalty is less relevant when using
64-bit words instead of 32-bit words.

When there are long runs of dirty words in some of the bitmaps,  EWAH  might
be preferable---it will access each dirty word at most once, whereas
a WAH decoder checks the first bit of each dirty word to ascertain it is
a dirty word. An EWAH decoder can skip a sequence of dirty words whereas a WAH decoder must access them all. For example, if we compute a logical AND between a bitmap
containing only dirty words, and another containing very few non-zero words, the running time of the operation with EWAH compression
will only depend on the small compressed size of the second bitmap.

\subsection{Constructing a bitmap index}

Given $L$~bitmaps and a table having $n$~rows and $c$ columns, 
we can na\"{\i}vely construct a bitmap index in
time $O(n L)$ by appending a word to each compressed bitmap  every 
32 or 64~rows.
We found this approach impractically slow when $L$ was large---typically, with
$k=1$.
Instead, we construct bitmap indexes in time 
proportional to the size of the index 
(see Algorithm~\ref{algo:owengenbitmap}): within each block of $w$~rows (e.g., $w=32$), we store the values of the bitmaps
in a set---omitting any unsolicited bitmap, whose values are all false (0x00). 
We use the fact
we can add several clean words of the
same type to a compressed bitmap in constant time.

Our implementation is able to generate the index efficiently on disk, even with extremely large tables and millions of (possibly small)
compressed bitmaps, using
\owen{there, I think I motivated it enough, without claiming it could not
be done some other way.}
horizontal partitioning:
we divide the table's rows into large blocks, such that each block's
compressed index fits in  a fixed memory 
budget (256\,MiB). Each block of bitmaps is written sequentially~\cite{jurgens2001tbi} and preceded
by an array of 4-byte integers containing the location of each bitmap within the block. 

\begin{algorithm}
\small
\begin{algorithmic}
\STATE Construct: $B_1,\ldots, B_L$, $L$ compressed bitmaps
\STATE $\mathrm{length}(B_i)$ is current (uncompressed) length (in bits) of bitmap $B_i$ 
\STATE $w$ is word length in bits, a power of 2 (e.g., $w = 32$)
\STATE $\omega_i \leftarrow 0$ for $1 \leq i \leq L$.
\STATE $c\leftarrow 1$ \COMMENT{row counter}
\STATE $\mathcal{N} \leftarrow \emptyset$ \COMMENT{$\mathcal{N}$ records the dirtied bitmaps}
\FOR{each table row}
\FOR{each attribute in the row}
\FOR {each bitmap  $i$ corresponding to the attribute value}
\STATE set to true the $(c \bmod w)^{\mathrm{th}}$~bit of word $\omega_i$
\STATE $\mathcal{N} \leftarrow \mathcal{N} \cup \{i\}$
\ENDFOR
\ENDFOR
\IF{$c$ is a multiple of $w$} 
\FOR{$i$ in $\mathcal{N}$}
\STATE add $c/w-\mathrm{length}(B_i)  - 1$~clean words  (0x00) to $B_i$ 
\STATE add the word $\omega_i$ to bitmap $B_i$
\STATE $\omega_i \leftarrow 0$
\ENDFOR
\STATE $\mathcal{N} \leftarrow \emptyset$
\ENDIF
\STATE  $c\leftarrow c+1$  
\ENDFOR
\FOR{$i$ in \{1,2,\ldots,L\}}
\STATE add $c/w-\vert B_i \vert  - 1$~clean words (0x00) to $B_i$ 
\ENDFOR
\end{algorithmic}
\caption{\label{algo:owengenbitmap}
Constructing bitmaps. For simplicity, we assume the number of
rows is a multiple of the word size.}
\end{algorithm}

\subsection{Faster operations over compressed bitmaps}

Beside compression, there is another reason to use
RLE: it makes operations faster~\cite{wu2006obi}. Given
(potentially many) 
 compressed bitmaps
 $B_1,\ldots,B_L$ of sizes $\vert B_i \vert$, 
 Algorithm~\ref{algo:genrunlengthmultiple} computes
 $\land_{i=1}^L B_i$ and  $ \lor_{i=1}^L B_i$
in time\footnote{Unless otherwise stated,
we use RLE compression with $w$-bit counters. In the complexity analysis, we do not bound the number of rows $n$. 
}
$O(L \sum_i |B_i|)$.
For BBC, WAH, EWAH and all similar RLE variants, similar 
algorithms exists: we only present the results for traditional RLE to
simplify the exposition.

Indeed, within a given pass through the main loop of 
Algorithm~\ref{algo:genrunlengthmultiple},
we need to compute the minimum and the maximum between 
$L$ $w$-bit counter values which requires $O(L)$~time.
Hence, the running time is determined by the number of iterations, which
is bounded by the sum of the compressed sizes of the bitmaps ($\sum_i |B_i|$).

For RLE with variable-length counters, the runs are encoded using $\log n$~bits and so each pass through the main loop of Algorithm~\ref{algo:genrunlengthmultiple} will be in $O(L \log n)$,
and a weaker result is true: the computation is in time
$O(L \sum_i |B_i| \log n)$. We should avoid concluding that the complexity is worse due to the $\log n$ factor: variable-length RLE can 
 generate smaller bitmaps than fixed-length RLE.

\begin{algorithm}
\small
\begin{algorithmic}
\STATE \textbf{INPUT:} $L$ bitmaps $B_1, \ldots B_L$ 
\STATE $I_i \leftarrow$ iterator over the runs of identical bits of  $B_i$
\STATE $\Gamma \leftarrow$ representing the aggregate of $B_1, \ldots B_L$  (initially empty)
\WHILE{some iterator has not reached the end}
\STATE let $a'$ be the maximum of all starting values for the runs of $I_1, \ldots, I_L$
\STATE let $a$ be the minimum of all ending values for the runs of $I_1, \ldots, I_L$
\STATE append run $[a',a]$ to $\Gamma$ with value determined by 
$\gamma(I_1, \ldots, I_L)$
\STATE increment all iterators whose current run ends at $a$.
\ENDWHILE
\end{algorithmic}
\caption{\label{algo:genrunlengthmultiple}Generic $O(L \sum_i |B_i| )$ algorithm to compute any 
 bitwise operations  between 
$L$~bitmaps. We assume the $L$-ary bitwise operation, $\gamma$, itself is in $O(L)$. 
}
\end{algorithm}

A stronger result is possible if the bitwise operation is updatable in $O(\log L)$ time.
That is, given     the result of an updatable $L$-ary operation $\gamma(b_1, b_2,\ldots,b_L)$, we can compute the updated value when a single bit is modified ($b'_i$),
\begin{eqnarray*}\gamma(b_1, b_2,\ldots,b_{i-1},b'_i,b_{i+1},\ldots, b_L),\end{eqnarray*} in $O(\log L)$ time.
All \emph{symmetric} Boolean functions are so updatable: we merely
maintain a count of the number of ones, which (for a symmetric function)
determines its value.    Symmetric functions include AND, OR, NAND, NOR,
XOR and so forth.
For example, given the number of 1-bits in a set of $L$~bits, we can update their
logical AND or logical OR aggregation ( $\land_{i=1}^L b_i$, $\lor_{i=1}^L b_i$) in
constant time given that one of the bits changes its value. 
Fast updates also exist for functions 
that are symmetric except that specified inputs
are inverted (e.g., Horn clauses).

From Algorithm~\ref{algo:genrunlengthmultiplefaster}, we have the following lemma. (The result is presented for fixed-length counters; when using variable-length counters, 
multiply the complexity by $\log n$.)

\begin{lemma}\label{rlelogical} Given $L$~RLE-compressed bitmaps
of sizes
$|B_1|, |B_2|, \ldots, |B_L|$ and any bitwise logical operation computable in $O(L)$ time, 
the aggregation of the bitmaps 
 is in time $O( \sum_{i=1}^L \vert B_i \vert L )$.
 If the bitwise operation is updatable in $O(\log L)$ time,    
 the
 aggregation is in time $O( \sum_{i=1}^L \vert B_i \vert  \log L)$. 
\end{lemma}

\begin{algorithm}
\small
\begin{algorithmic}
\STATE \textbf{INPUT:} $L$ bitmaps $B_1, \ldots B_L$ 
\STATE $I_i \leftarrow$ iterator over the runs of identical bits of  $B_i$
\STATE $\Gamma \leftarrow$ representing the aggregate of   $B_1, \ldots B_L$  (initially empty)
\STATE $\gamma$ be the bit value determined by $\gamma(I_i,\ldots, I_L)$
\STATE $H'$ is an $L$-element max-heap storing starting values of the runs (one per bitmap)
\STATE $H$ is an $L$-element min-heap storing ending values of the runs and an indicator of which bitmap
\STATE 
        a table $T$ mapping each bitmap to its entry in $H'$
\WHILE{some iterator has not reached the end}
\STATE let $a'$ be the maximum of all starting values for the runs of $I_1, \ldots, I_L$, determined from $H'$
\STATE let $a$ be the minimum of all ending values for the runs of $I_1, \ldots, I_L$, determined from $H$
\STATE append run $[a',a]$ to $\Gamma$ with value $\gamma$
\FOR {iterator $I_i$  with a run ending at $a$ (selected from $H$)}
\STATE increment $I_i$ while updating $\gamma$ in $O(\log L)$ time
\STATE pop $a$ value from $H$, insert new ending run value to $H$
\STATE from hash table, find old starting value in $H'$, and increase it to the new starting value
\ENDFOR
\ENDWHILE
\end{algorithmic}
\caption{\label{algo:genrunlengthmultiplefaster}Generic $O(\sum_i |B_i| \log L )$ algorithm to compute any bitwise operations  between 
$L$~bitmaps  updatable in $O(\log L)$ time. 
}
\end{algorithm}

\begin{corollary} 
This result is also  true for word-aligned (BBC, WAH or EWAH) compression.
\end{corollary}



\begin{figure}
\includegraphics[width=\textwidth]{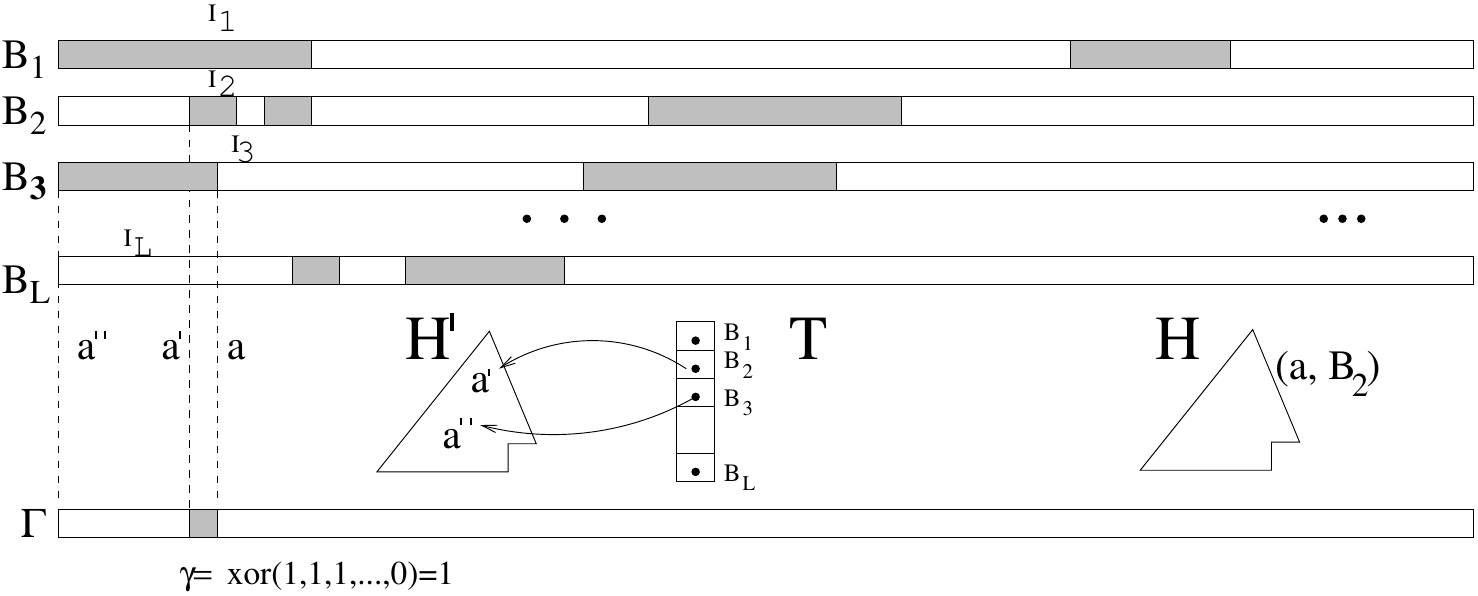}
\caption{\label{fig:visualizeXor}Algorithm~\ref{algo:genrunlengthmultiplefaster} in action. }
\end{figure}

See Fig.~\ref{fig:visualizeXor}, where we show the XOR of $L$ bitmaps.
This situation depicted has just had $I_2$ incremented, and $\gamma$ is
about to be updated to reflect the change of $B_2$ from ones to zeros.
The value of $a$ will then be popped from $H$, whose minimum value will
then be the end of the $I_1$ run.  Table $T$ will then allow us to find and 
increase the key of $B_2$'s entry in $H'$, where it will become $a+1$
and likely be promoted toward the top of $H'$.

In the rest of this section, we assume an RLE encoding
such that the merger of two running lengths reduces the total size
(0 repeated $x$~times
and 0 repeated $y$~times, becomes 0 repeated $x+y$~times). These encodings include
BBC, WAH and EWAH.  
We also consider only fixed-length counters; for variable-length
counters, the running time complexity should have the
bitmap index size multiplied by $\log n$.
\owen{tweaked the wording.}

From Algorithm~\ref{algo:genrunlengthmultiplefaster}, we have
that $| \land_{i\in S} B_i | \leq | \sum_{i\in S} B_i | $,
$| \lor_{i\in S} B_i | \leq | \sum_{i\in S} B_i | $, and so on for other binary bitwise operation such as $\oplus$. 
This bound is practically optimal: e.g., the logical AND of 
the bitmaps 10{\ldots}10 ($n$~runs)
and 11{\ldots}11 (1~run)  is  10{\ldots}10 ($n$~runs).
\killforgood{For bitmaps with few 1-bits,  
we expect $| B_1 \land B_2 |\leq  \max(| B_1 | , | B_2 |)$ to hold
approximately,
and similarly, for bitmaps with few 0-bits, 
we expect $| B_1 \lor B_2 | \leq  \max(| B_1 | , | B_2 |)$.
}

Hence, for example, when computing $B_1 \land B_2 \land B_3\land +\cdots\land B_L$ we may start with the computation of
$B_1 \land B_2=B_{1,2}$ in $O(|B1|+|B2|)$ time. The bitmap $B_{1,2}$ is of
size at most $|B_1|+|B_2|$, hence $B_{1,2} \land B_3 $ can be done in time $O(|B_1|+|B_2|+|B_3|)$. Hence, the total running time is in $O(\sum_{i=1}^L (L-i+1) |B_i|)$.

Hence, there are at least three different generic algorithms to aggregate a set of
$L$~bitmaps for these most common bitwise operations:
\begin{itemize}
\item We use Algorithm~\ref{algo:genrunlengthmultiplefaster}, which runs in time $O( (\sum_{i=1}^L  |B_i|) \log L)$.
It generates a single output bitmap, but it uses two L-element heaps.
It works for a wide range of queries, not only simple queries such as $\lor_{i=1}^L B_i$.
\item  
We aggregate  two bitmaps at
a time starting with $B_1$ and $B_2$, then aggregating the result with $B_3$, and so on. This
requires  
 time $O( \sum_{i=1}^L (L-i+1) |B_i|)$. While only a single temporary compressed bitmap is held in
 memory, $L-1$~temporary bitmaps are created. To minimize processing time, the input bitmaps
 can be sorted in increasing size.
 \item We can store the bitmaps in a priority queue~\cite{1316694}. We repeatedly pop the two
 smallest bitmaps, and insert the aggregate of the two bitmaps. This approach runs in time
 $O( (\sum_{i=1}^L  |B_i|) \log L)$, and it generates $L-1$~intermediate bitmaps. 
\item Another approach is to use in-place computation~\cite{1316694}: (1)~an uncompressed
bitmap is created in time $O(n)$ (2)~we aggregate the uncompressed
bitmap with the each one of the compressed bitmaps (3)~the row IDs are extracted
from the uncompressed bitmap in time $O(n)$. For logical OR (resp.\ AND) aggregates,
the uncompressed bitmap is initialized with zeroes  (resp. ones).
The total cost is in $O( L n )$: 
$L$ passes over the uncompressed bitmap will be
required. However, when processing each compressed bitmap, we can skip over
portions of the uncompressed bitmaps e.g., when we compute a logical OR, we can omit runs of zeroes.  
If the table has been horizontally partitioned, it will be
possible to place the uncompressed bitmap in main memory.
\end{itemize}
We can minimize the complexity by choosing the algorithm after
loading the bitmaps. 
For example,  to compute logical OR over many bitmaps with long runs of zeroes---or logical AND over many bitmaps with long runs of ones---an in-place computation might be preferable. When there are few bitmaps, computing the operation two bitmaps at
a time is probably efficient. Otherwise, using Algorithm~\ref{algo:genrunlengthmultiplefaster} or a priority queue~\cite{1316694} might be advantageous. Unlike the alternatives,
Algorithm~\ref{algo:genrunlengthmultiplefaster} is not limited to simple  queries such as $\lor_{i=1}^L B_i$.

\section{Finding the best reordering is NP-Hard}
\label{sec:FindingthebestreorderingisNPHard}

Let $d(r,s)$ be the number of bits differing between rows $r$ and $s$.
Our problem is to find the best ordering of the rows $r_i$ so as
to minimize $\sum_i d(r_i,r_{i+1})$.
Pinar et al.\ have 
reduced the row-reordering problem to the 
Traveling Salesman Problem 
(TSP)~\cite[Theorem 1]{pinar05} using $d$
as the distance measure. 
Because $d$
satisfies the triangle
inequality,  the row-reordering problem can be approximated
with 1.5-optimal cubic-time algorithms~\cite{Christofides1976}.
Pinar and Heath~\cite{331562} proved that
the row-reordering problem is NP-Hard by reduction from the Hamiltonian path problem.

\cutmarch{
\begin{theorem}\label{thm:trivialTCS}
The bitmap-index row-reordering problem is NP-Hard.
\end{theorem}
\begin{pf}
The Hamiltonian path problem is the problem of finding a path
that visits every vertex exactly once. It is NP-Complete. 
In a similar manner as Johnson et al.~\cite{johnson2004clb}, we reduce
the Hamiltonian path problem  to the row-reordering problem. Consider
any graph $G$. Construct the matrix $M$ as follows: each column
of the matrix represents a vertex and each row an edge. 
For each row $r$ representing edge $(v,w)$, set $M_{r,v}=M_{r,w}=1$.
All other entries are zero.
When two edges $r$ and $s$ have a vertex in common, then $d(r,s)=1$,
otherwise $d(r,s)=2$. Hence, if we minimize  $\sum_i d(r_i,r_{i+1})$,
we find an Hamiltonian path if there is one.
\end{pf}
}

However, 
the hardness of the problem depends on $L$ being variable.
If the number $L$ of bitmaps  were a constant,  
the next lemma shows that 
the problem 
would
\emph{not} 
be
NP-hard\footnote{Assuming P $\not =$ NP.}: 
an (impractical)
linear-time solution is possible.

\begin{lemma}\label{prop:OwenPolyTime}
For any constant number of bitmaps $L$, the row-reordering
problem requires only 
$O(n)$ time.
\end{lemma}
\begin{pf}
Suppose that an optimal row ordering is such that
identical rows do not appear consecutively.
Pick any row value---any sequence of $L$~bits appearing
in the bitmap index---and call it $a$. Consider two
occurrences of $a$, where one occurrence of the row value $a$ appears
 between the row values $b$ and $c$: we may have $b=a$ and/or $c=a$. 
Because the Hamming
distance satisfies the triangle inequality, we have
$d(b,c) \geq d(b,a) + d(a,c)$. Hence, we can move the occurrence
of $a$ from between $b$ and $c$, placing it instead
with any other occurrence of $a$---without increasing total cost,  $\sum_i d(r_i,r_{i+1})$.
Therefore, there is an optimal solution with all identical
rows clustered.

In a bitmap index with $L$~bitmaps, there are 
only $2^L$ different possible distinct rows,
irrespective of the total number of rows $n$.
Hence, there are at most $(2^L)!$~solutions to enumerate
where all identical rows are clustered,
which concludes the proof.
\end{pf}



If we generalize the row-reordering problem to the word-aligned case,
the problem is still NP-hard.
We can formalize the problem as such: \suggestwecut{you must} order the rows
in a bitmap index such that the storage cost of any sequence
of identical clean words (0x00 or 1x11) costs $w$~bits whereas the
 cost of any other word is $w$~bits.
\begin{theorem}
The word-aligned row-reordering problem is NP-hard if the number of bits per word ($w$) is a constant.
\end{theorem}
\begin{pf}
Consider the case where each row of the bitmap is repeated $w$~times.
 It is possible to reorder these identical rows so that they form only clean words (1x11 and 0x00). 
 There exists an optimal
 solution to the word-aligned row-reordering problem obtained by
 reordering these blocks of $w$ identical rows.
 The problem of reordering these clean words is equivalent to the
 row-ordering problem, which is known to be NP-hard. 
\end{pf}


\section{Sorting to improve compression}
\label{sec:SortingtoImproveCompression}

Sorting can benefit bitmap indexes at several levels.
We can sort the rows of the table.
The sorting order depends itself on the order of the table columns.
And finally, we can allocate the bitmaps to the attribute values in sorted order.

\subsection{Sorting rows}
\label{sec:sorting-rows}

Reordering the rows of a compressed bitmap index can improve
compression. Whether 
 using RLE, BBC, WAH or EWAH,
the problem is NP-hard 
(see \S~\ref{sec:FindingthebestreorderingisNPHard}). 
A simple heuristic begins with an uncompressed
index.  Rows (binary vectors)
are then rearranged to promote runs.  In the process, we may also
reorder the bitmaps.
%
This is the approach of Pinar et al.~\cite{pinar05}, Sharma and Goyal~\cite{sharma2008emc}, Canahuate et al.~\cite{pinarunpublished}, and
Apaydin et al.~\cite{apay:data-reordering},
but it uses $\Omega(n L)$ time.
For the large dimensions and number of rows we have considered,
it is infeasible. 
A more practical approach 
is to reorder the  table,
then construct the compressed index directly (see \S~\ref{subsubsection:Gray-Lexallocation}); we can also reorder the 
table columns prior to sorting (see \S~\ref{sec:dimOrder}).


Sorting lexicographically large files in external memory is not excessively 
expensive~\cite{1132964,yiannis2007ctf}. With a memory buffer of $M$~elements, we can
sort almost $M^2$~elements in two passes. \cutmarch{ Consider the problem \owen{hates ``you'' in this writing:  where you must sort: propose instead} of sorting $n$ elements with a memory buffer of only $M$~elements.
The algorithm used by the Unix
 \textbf{sort} command 
works in two steps~\cite{yiannis2007ctf}~: }
\cutmarch{
\begin{enumerate}
\item  $\lceil n/M \rceil$~non-overlapping blocks
of $M$~elements are read, sorted and written on disk, one by one.
\item Part of the memory ($x$~elements) is allocated 
as a disk buffer whereas the rest of the memory is
used to read the first $z=\lfloor (M-x)/\lceil n/M \rceil \rfloor$~elements of each block. The elements read are stored in a heap. The
smallest element of the heap is removed and placed in the disk buffer. We repeat this operation; as soon as one of the $\lceil n/M \rceil$~input buffers is empty, it is replenished with the next  $z$ elements of the block. As soon as the disk buffer is full, it is written to disk.
\end{enumerate}
Hence, with a memory buffer of $M$~elements, we can
sort almost $M^2$~elements in two passes, without any random write access to disk. A computer with a few gigabytes of RAM can sort one or two terabytes of data. If RAM is limited, there are variants of this algorithm using  $\lceil \log_{M} n \rceil$~passes.
}

Several types of ordering can be used for ordering rows.
\begin{itemize}
\item  In lexicographic order, a sequence
$a_1, a_2, \ldots$ is smaller than another sequence $b_1, b_2, \ldots$ if
and only if there is a $j$ such that $a_j < b_j$ and $a_i = b_i$ for $i<j$.
The Unix \textbf{sort} command provides an efficient
means of sorting flat files into lexicographic order; in under 10\,s
our test computer (see \S~\ref{sec:Experiment})  sorted a
5-million-line,
120\,MB file. SQL supports lexicographic sort 
via ORDER BY. 
\item We may cluster runs of identical  rows. This problem can be solved
with hashing algorithms, by multiset discrimination algorithms~\cite{cai1995umd}, or by a
lexicographic sort.
While sorting requires $\Omega(n \log n)$ time, clustering
identical facts requires only linear time ($O(n)$).
However,  
the relative efficiency of clustering decreases drastically with the number of dimensions. 
The reason is best illustrated with an example. Consider lexicographically-sorted
 tuples  $(a,a)$, 
$(a,b)$,
$(b,c)$,
$(b,d)$. Even though all these tuples are distinct, the lexicographical order is 
beneficial to the first dimension. Random multidimensional row clustering fails to cluster the values within columns.
\item Instead of fully ordering all of the rows, we may reorder rows only within disjoint blocks (see \S~\ref{subsection:crazyblocks}). Block-wise sorting is not competitive.
\item Gray-code (GC) sorting, examined next.
\end{itemize}

GC sorting
 is defined over bit vectors~\cite{pinar05}. The 
 list of 2-of-4 codes in increasing order is 0011, 0110, 0101, 1100, 1010, 1001. Intuitively, the further right the first bit is, the smaller the code is, just as in the lexicographic order. However, contrary to the lexicographic order,  the further left the second bit is, the smaller the code is. Similarly, for a smaller code, the third bit should be further right, the fourth bit should be further left and so on. Formally, we define the Gray-code order as follows.

\begin{definition}{\label{GCordering}}
The sequence $a_1,a_2, \ldots$ is smaller than  
$b_1,b_2,\ldots $ if and only if there exists $j$ such that\footnote{The symbol $\lxor$ is the XOR operator.} 
$a_j = a_1 \lxor a_2 \lxor \ldots \lxor a_{j-1}$,
$b_j \not = a_j$, and $a_i=b_i$ for $i<j$.
\end{definition}

We denote this ordering by $<_{\mathrm{gc}}$, as opposed to the normal
lexicographic ordering, $<_{\mathrm{lex}}$. The reflexive versions of 
these are $\leq_{\mathrm{gc}}$ and $\leq_{\mathrm{lex}}$, respectively.
 
Algorithm~\ref{algo:graycomp}
, an adaptation of
Ernvall's procedure~\cite{ernvall1984csp,richards1986dca} to sparse data,
 shows how to compare sparse GC bit vectors
$v_1$ and $v_2$ in time $O(\min(\vert v_1 \vert,\vert v_2 \vert)$ where
$ \vert v_i\vert$ is the number of true value in bit vector $v_i$.
Sorting the rows of a bitmap index without materializing the uncompressed
bitmap index is possible
: 
we implemented an $O(n c k \log n)$-time solution
for $k$-of-$N$ indexes using
an external-memory B-tree~\cite{qdbm} ($c$ is the number of columns). As values, we used the rows of the table, and as keys, we used
the position of the ones in the bitmap row as 32-bit integers---some of our indexes
have half a million bitmaps. 
Hence, we used $4ck$~bytes per row for storing the keys
alone.
Both keys and values were compressed using LZ77 to minimize I/O costs---compression
improved performance noticeably in our informal tests.
We expected that this implementation would be significantly slower
than lexicographic sorting, but the degree of difference surprised us:
%
our implementation proved to be \emph{two orders of  
magnitude} slower than lexicographic sort using the Unix \textbf{sort} command.

For some of our tests (see \S~\ref{sec:bitmap-reordering-exper}), we wish to rearrange the order of the bitmaps
prior to GC sorting. We get this result  
by applying the appropriate permutation to the positions that
form the B-tree keys, during the B-tree's construction.

\begin{algorithm}
\begin{algorithmic}
\STATE \textbf{INPUT}: arrays $a$ and $b$ representing the position of the ones in two bit vectors, $a'$ and $b'$
\STATE \textbf{OUTPUT}: whether $a' <_{\mathrm{gc}} b'$ 
\STATE $f\leftarrow \texttt{true}$
\STATE $m \leftarrow \min(\textrm{length}(a), \textrm{length}(b))$
\FOR{$p$ in $1,2,\ldots,m$}
\STATE return $ f$ if $a_p > b_p$ and $ \lnot f$ if $a_p < b_p$
 \STATE $f\leftarrow \lnot f$
 \ENDFOR
 \STATE return $\lnot f$ if $\textrm{length}(a) > \textrm{length}(b)$,
 $f$ if $\textrm{length}(b)>\textrm{length}(a)$, and  \texttt{false} otherwise
\end{algorithmic}
\caption{\label{algo:graycomp}Gray-code less comparator between  sparse bit vectors}
\end{algorithm}

For RLE, the best ordering of the rows
of a bitmap index minimizes the sum of the Hamming distances:
$\sum_i h(r_i,r_{i+1})$ where $r_i$ is the $i^{\mathrm{th}}$~row, for
$h(x,y) = \left | \rule{0mm}{0.95em} \{i | x_i\not= y_i\}\right |$.
If all $2^L$~different rows are present, the GC sort would
be an optimal solution to this problem~\cite{pinar05}. The following
proposition shows that GC sort is also optimal
if all ${N \choose k}$ $k$-of-$N$ codes are present. The same is 
false of lexicographic order when $k>1$: 0110 immediately follows 1001 among
2-of-4 codes, but their Hamming distance is 4.

\begin{proposition}\label{prop:graycode}We can enumerate,  in GC order, all $k$-of-$N$ codes
in time $O(k {N \choose k})$ (optimal complexity).
Moreover, the Hamming distance between successive codes is minimal (=2).
\end{proposition}
\begin{pf}
Let $a$ be an array of size $k$ indicating the positions of the ones in $k$-of-$N$ codes.
As the external loop, vary the value $a_1$ from 1 to $N-k+1$. Within this loop,
vary the value $a_2$ from $N-k+2$ down to $a_1+1$. Inside this second loop,
vary the value of $a_3$ from $a_2+1$ up to $N-k+3$, and so on.
By inspection, we see that all possible  codes are generated in decreasing GC order.
%
To see that the Hamming distance between successive codes is 2, consider what
happens when $a_i$ completes a loop. Suppose that $i$ is odd and greater than 1,
 then $a_i$ 
had value $N-k+i$ and it will take value $a_{i-1}+1$. Meanwhile, by construction,
 $a_{i+1}$ (if
it exists) remains at value $N-k+i+1$ whereas $a_{i+2}$ remains at value
$N-k+i+2$ and so on. The argument is similar if $i$ is even.
\end{pf}


For encodings like BBC, WAH or EWAH, GC sorting is suboptimal,
even when all $k$-of-$N$ codes are present. For example consider the
sequence of rows 1001, 1010, 1100, 0101, 0101, 0110, 0110, 0011. Using 4-bit
words, we see that a single bitmap contains a clean word (0000)
whereas by exchanging the fifth and second row, we get two clean words (0000
and 1111).

\subsection{Sorting bitmap codes}
\label{subsection:Sortingbitmapcodes}
For a simple 
index, the map from  attribute values to bitmaps is
inconsequential; for $k$-of-$N$ encodings, some bitmap allocations are more
compressible: consider an attribute with
two overwhelmingly frequent values and many other values that
occur once each.  If the table rows are given in random order,
the two frequent values should have codes that
differ in Hamming distance 
as little 
as possible to maximize compression (see Fig.~\ref{fig:trivialHammingresult} for an example).  However, it is also important to allocate bitmaps well when
the table is sorted, rather than randomly ordered. 
\owen{It seemed there was a conceptual break between this para and the next,
so I added the previous sentence.}
\begin{figure}
\centering
\subfigure{
\begin{tabular}{cccc}
1& 0 & 0 & 1\\
0& 1 & 1 & 0\\
1& 0 & 0 & 1\\
0& 1 & 1 & 0\\
0& 1 & 1 & 0\\
1& 0 & 0 & 1
\end{tabular}
}
\subfigure{
\begin{tabular}{cccc}
1& 0 & \textbf{0} & 1\\
1& 1 & \textbf{0} & 0\\
1& 0 & \textbf{0} & 1\\
1& 1 & \textbf{0} & 0\\
1& 1 & \textbf{0} & 0\\
1& 0 & \textbf{0} & 1
\end{tabular}
}
\caption{Two bitmaps representing the sequence of values a,b,a,b,b,a using different codes. If codes have a Hamming distance of two (right), the result is more compressible than if the Hamming distance is four (left).}
\label{fig:trivialHammingresult}
\end{figure}

There are several ways to allocate the bitmaps.
Firstly, the attribute values can be visited in alphabetical or numerical order,
or---for histogram-aware schemes---in order of frequency. Secondly, the bitmap codes
can be used in different orders. We consider lexicographical ordering
 (1100, 1010, 1001, 0110, \ldots) and GC
order (1001, 1010, 1100, 0101, \ldots) ordering (see proof of Proposition~\ref{prop:graycode}). 

\emph{Binary-Lex} denotes sorting the table lexicographically 
and allocating bitmap codes so that
the $i^{\mathrm{th}}$ attribute gets the $i^{\mathrm{th}}$
numerically smallest bitmap code, when codes are viewed as binary numbers.  
\emph{Gray-Lex} is similar, except that 
the $i^{\mathrm{th}}$ attribute gets the rank-$i$ bitmap code
in GC order.  (Binary-Lex and Gray-Lex coincide when $k=1$.)
These two approaches are histogram oblivious---they
ignore the frequencies of attribute values.



Knowing the frequency 
of each attribute value 
can improve code assignment when $k>1$. 
\suggestwecut{For instance,
clustering dirty words increases the compressibility.}%
Within a column, {Binary-Lex} and {Gray-Lex} order runs of identical values
irrespective of the frequency: the sequence 
\texttt{afcccadeaceabe} may  become 
\texttt{aaaabccccdeeef}. For better compression, we should order
the attribute values---within a column---by their frequency 
(e.g., \texttt{aaaacccceeebdf}). Allocating the bitmap codes in
GC order to the frequency-sorted attribute values, our \emph{Gray-Frequency}
sorts the table rows as follows.
Let $f(a_i)$ be
the frequency  of attribute $a_i$. Instead
of sorting the table rows $a_1, a_2,\ldots, a_d$, we
lexicographically sort the extended rows
$f(a_1), a_1, f(a_2), a_2,\ldots, f(a_d), a_d$
(comparing the frequencies numerically.) 
The frequencies $f(a_i)$ are 
discarded prior to indexing.

\subsubsection{No optimal ordering when $k>1$}

\emph{No} allocation scheme is optimal for all tables,
even if we consider only lexicographically sorted tables. 

\begin{proposition}\label{prop:all-allocs-bad}
For any allocation $\mathcal{C}$ of attribute values to $k$-of-$N$ codes, there is a table
where $\mathcal{C}$ leads to a suboptimal index. 
\end{proposition}

\begin{pf}
Consider a lexicographically sorted table, where we encode the 
second column with 
$\mathcal{C}$. 
 We construct a table
where $\mathcal{C}$ is worse than some other ordering $\mathcal{C}'$.
The first column of the table is for attribute $A_1$, which is the
primary sort key, and the second column is for attribute $A_2$.  Choose any two
attribute values $v_1$ and $v_2$ from $A_2$, where
$\mathcal C$ assigns codes of maximum Hamming distance (say $d$)
 from one another.  If $A_2$ is large enough, $d>2$.
Our bad input table has unique ascending values in the first column, 
and the second column alternates between $v_1$ and $v_2$.
Let this continue for $w$ rows.
On this input, there will be $d$ bitmaps that are entirely dirty for
the second column\footnote{
There are other values in $A_2$ and if we must use them, let them occur
once each, at the end of the table, and make a table whose length is a large
multiple of $w$.}. Other bitmaps in the second column are made entirely of identical clean words.

Now consider $\mathcal{C}'$, some allocation that assigns $v_1$
and $v_2$ codewords at Hamming distance 2.   On this input, $\mathcal{C'}$ produces 
only 2 dirty words in the bitmaps for $A_2$.  This is fewer dirty words
than $\mathcal{C}$ produced.  

Because bitmaps containing only identical clean words use less
storage than bitmaps made entirely of dirty words, we have that
allocation $\mathcal{C}'$ will compress the second column better.
This concludes the proof. 
\end{pf}


\subsubsection{Gray-Lex allocation and GC-ordered indexes}
\label{subsubsection:Gray-Lexallocation}
Despite the pessimistic result of Proposition~\ref{prop:all-allocs-bad}, we can
focus in choosing good allocations for special cases, such as dense indexes
(
including those where most 
of the 
 possible rows appear
),  or for typical sets of data.

For dense indexes, GC sorting is better~\cite{pinar05} at minimizing
the number of runs, a helpful effect even with word-aligned schemes.
However, as we already pointed out, the approach used by  Pinar et al.~\cite{pinar05} requires $\Omega(n L)$ time. For technical reasons,
 even our more economical B-tree approach is much slower than lexicographic sorting. As an alternative, we propose
a low-cost way to GC sort $k$-of-$N$ indexes, using only lexicographic
sorting and Gray-Lex allocation.

We now examine  Gray-Lex allocation more carefully,  to prove
that its results are equivalent to building the uncompressed index,
GC sorting, and then  compressing the index.


Let $\gamma_i$ be the invertible mapping from attribute $i$ to the $k_i$-of-$N_i$
code---written as an $N_i$-bit vector. Gray-Lex implies a form of monotonicity: for $a$ and $a'$
belonging to the $i^{\mathrm{th}}$ attribute, $A_i$, 
$a \leq 
 a' \Rightarrow 
\gamma_i(a) \leq_{\mathrm{gc}} \gamma_i(a')$.  
The overall encoding of a table row $r = (a_1, a_2, \ldots, a_c)$ is 
obtained by applying each $\gamma_i$  to $a_i$, and concatenating
the $c$ results.  I.e., $r$ is encoded into
$$ \Gamma(r) = (\overbrace{\alpha_1, \alpha_2, \ldots \alpha_{N_1}}^{\gamma_1(a_1)},
    \overbrace{\alpha_{N_1+1}, \ldots \alpha_{N_1+N_2}}^{\gamma_2(a_2)},
    \ldots
    \overbrace{\alpha_{L-N_c+1}, \ldots \alpha_L}^{\gamma_c(a_c)})
$$
where  $\alpha_i\in \{0,1\}$ for all $i$.

First, let us assume that we use only $k$-of-$N$ codes, for $k$ even. Then, the following proposition holds.
    
\begin{proposition}\label{k-even-gclex-works}
Given two table rows $r$ and $r'$, using Gray-Lex $k$-of-$N$ codes  for $k$ even, we have
$r \leq_{\mathrm{lex}} r' \iff \Gamma(r) \leq_{\mathrm{gc}} \Gamma(r')$. The values of $k$ and $N$ can vary from column to column.
\end{proposition}

\begin{pf}
We write $r = (a_1, \ldots , a_c)$ and $r' = (a'_1, \ldots , a'_c)$. We note $(\alpha_1,\ldots,\alpha_L) = \Gamma(r)$ and
$(\alpha'_1,\ldots,\alpha'_L) = \Gamma(r')$.
Without loss of generality, we assume $\Gamma(r) \leq_{\mathrm{gc}} \Gamma(r')$. First, if $\Gamma(r) = \Gamma(r')$, then $r = r'$ since each $\gamma_i$ is
invertible.

We now proceed to the case where $\Gamma(r) <_{\mathrm{gc}} \Gamma(r')$.
Since they are not equal, Definition~\ref{GCordering} implies there is
a bit position $t$ where they first differ; at position $t$, we have that
$\alpha_t = \alpha_1 \oplus \alpha_2 \oplus \cdots \oplus \alpha_{t-1}$.
Let $\hat t$ denote the index of the attribute associated with bitmap $t$.
In other words, 
$N_1 + N_2 + \cdots + N_{\hat t - 1} < t \leq N_1 + N_2 + \cdots + N_{\hat t}$.
Let $t'$ be the first bitmap of the block for attribute $\hat t$; i.e.,
$t' = N_1 + N_2 + \cdots + N_{\hat t - 1}+1$.


$$
\begin{array}{rll}
  \Gamma(r) <_{\mathrm{gc}} & \Gamma(r') &\\
  \iff & \alpha_t = \bigoplus_{i=1}^{t-1} \alpha_i 
                                                     \ \wedge\  \alpha_t \not = \alpha'_t \ \wedge\  \bigwedge_{i=1}^{t-1}(\alpha_i = \alpha'_i) & \mbox{Def.~\ref{GCordering}}\\
    \iff & \alpha_t = \bigoplus_{i=1}^{t'-1} \alpha_i \oplus \bigoplus_{i=t'}^{t-1} \alpha_i 
           \ \wedge\  \alpha_t \not = \alpha'_t & \\
         &   \wedge\  \bigwedge_{i=1}^{t'-1}(\alpha_i = \alpha'_i) \ \wedge\  \bigwedge_{i=t'}^{t-1}(\alpha_i = \alpha'_i) & \mbox{associativity}\\
    \iff &  \alpha_t = 0 \oplus \bigoplus_{i=t'}^{t-1} \alpha_i 
         \   \wedge\  \alpha_t \not = \alpha'_t & \mbox{all codes are $k$-of-$N$}\\
         &  \wedge\  \bigwedge_{i=1}^{t'-1}(\alpha_i = \alpha'_i) \ \wedge\  \bigwedge_{i=t'}^{t-1}(\alpha_i = \alpha'_i) & \mbox{and $k$ is  even}\\
    \iff &  \alpha_t = \bigoplus_{i=t'}^{t-1} \alpha_i 
         \   \wedge\  \alpha_t \not = \alpha'_t & \\
         &  \wedge\  \bigwedge_{i=1}^{\hat t -1}(a_i = a'_i)\  \wedge \ \bigwedge_{i=t'}^{t-1}(\alpha_i = \alpha'_i) & \mbox{$\gamma_i$ is invertible}\\
    \iff &  \gamma_{\hat t}(a_{\hat t}) <_{\mathrm{gc}} \gamma_{\hat t}(a'_{\hat t}) 
         \   \wedge\  \bigwedge_{i=1}^{\hat t -1}(a_i = a'_i) & \mbox{Def.~\ref{GCordering}}\\
    \iff &  a_{\hat t} <_{\mathrm{lex}}\killforgood{^{\hat t} : Daniel
apparently thinks it is obvious to go from a full lex ordering to
the individual implied ordering on the components in the tuple; Owen
guesses he is probably right} a'_{\hat t}\  \wedge\  \bigwedge_{i=1}^{\hat t -1}(a_i = a'_i) & \mbox{$\gamma$s are monotone}\\
    \iff & r <_{\mathrm{lex}} r'  & \mbox{Def.\ of lex.\ order}\\
\end{array}
$$
\qed

\end{pf}

If some columns have $k_i$-of-$N_i$ codes with $k_i$ odd, then we have to reverse the order of the Gray-Lex allocation for some columns. Define the \emph{alternating Gray-Lex} allocation to be such that it has the Gray-Lex monotonicity    
($a \leq  a' \Rightarrow 
\gamma_i(a) \leq_{\mathrm{gc}} \gamma_i(a')$) when $\sum_{j=1}^{i-1} k_j$ is even, and is reversed ($a \leq  a' \Rightarrow 
\gamma_i(a) \geq_{\mathrm{gc}} \gamma_i(a')$) otherwise. Then we have the following lemma.

\begin{lemma}Given a table to be indexed with alternating Gray-Lex $k$-of-$N$ encoding, the following algorithms have the same output:
\begin{itemize}
\item Construct the bitmap index and sort bit vector rows using GC order.
\item Sort the table lexicographically and then construct the index.
\end{itemize}
The values of $k$ and $N$ can vary from column to column.
\end{lemma}

This result applies to any encoding where there is a fixed  number of 1-bits per column. Indeed, in these cases,
we are merely using  a subset of the $k$-of-$N$ codes.
For example, it also works with multi-component encoding where each
component is indexed using a unary encoding.

\subsubsection{Other Gray codes}

In addition to the usual Gray code, many other binary codes have the property
that any codeword is at Hamming distance 1 from its successor.  Thus, they can
be considered ``Gray codes'' as well, although we shall qualify them to
avoid confusion from our standard (``reflected'') Gray code. 

Trivially, we could permute columns in the Gray code table, or invert the
bit values in particular columns
(see Fig.~\ref{fig:invert-and-permute}).
However, there are other codes
that cannot be trivially derived from the standard Gray code.
Knuth~\cite[\S~7.2.1.1]{knut:vfour-fascicle-two} presents many
results for such codes.

\begin{figure}
\subfigure[3-bit reflected GC]{
\begin{minipage}{.3\textwidth}
\begin{tabular}{rrr}
\ \ \ \ \ 0&0&0\\
0&0&1\\
0&1&1\\
0&1&0\\
1&1&0\\
1&1&1\\
1&0&1\\
1&0&0\\
\end{tabular}
\end{minipage}
}
\subfigure[swap columns 1\&3]{
\begin{minipage}{.3\textwidth}
\begin{tabular}{rrr}
\ \ \ \ \ 0&0&0\\
1&0&0\\
1&1&0\\
0&1&0\\
0&1&1\\
1&1&1\\
1&0&1\\
0&0&1\\
\end{tabular}
\end{minipage}
}
\subfigure[then invert column 3]{
\begin{minipage}{.3\textwidth}
\begin{tabular}{rrr}
\ \ \ \ \ 0&0&1\\
1&0&1\\
1&1&1\\
0&1&1\\
0&1&0\\
1&1&0\\
1&0&0\\
0&0&0\\
\end{tabular}
\end{minipage}
}
\caption{\label{fig:invert-and-permute} The 3-bit reflected GC and two
other Gray codes obtained from it, first by exchanging the outermost
columns, then by inverting the bits in the third column.}
\end{figure}

   For us, three properties are important:
\begin{enumerate}
\item  Whether successive $k$-of-$N$ codewords have a Hamming distance of 2.
\item Whether the final codeword is at Hamming distance 1 from the 
initial codeword.  Similarly, whether the initial and final $k$-of-$N$ codewords
are at Hamming distance 2. 
\item Whether a collection of more than 2 successive codes (or more than
2 successive $k$-of-$N$ codes) has a small expected ``collective Hamming
distance''. 
(Count 1 for every bit position where at least two codes disagree. 
)
\end{enumerate}

The first property is 
important if we are assigning $k$-of-$N$
codes to attribute values.

The second property distinguishes, in Knuth's terminology, ``Gray paths'' from ``Gray cycles.''
It is important unless an attribute is the
primary sort key. 
E.g.,  the second column of a sorted table will have its values cycle from the smallest to the largest, again and again.

The third property is related to the ``long runs'' 
property~\cite{knut:vfour-fascicle-two,godd:long-run-codes} of
some Gray codes. 
Ideally, we would want to have long runs of identical values
when enumerating all codes.
However, for any $L$-bit Gray cycle, every code
word terminates precisely one run, hence the number of runs
is always $2^L$. Therefore, the average run length is always $L$.
The distribution of run lengths varies by code, however. 
 When $L$ is large, Goddyn and Grozdjak
show there are codes where
no run is shorter than $L - 3 \log_2 L$; in particular, for
$L=1024$, there is a code with no run shorter than 
1000~\cite{knut:vfour-fascicle-two,godd:long-run-codes}.
In our context, this property may  be
unhelpful: with $k$-of-$N$ encodings, we are interested in only those codewords of
Hamming weight $k$. 
Also, rather than have all
runs of approximately length $L$, we might prefer a few very long
runs (at the cost of many short ones).


One notable Gray code is
constructed by Savage and Winkler~\cite{sava:monotone-gray-codes},
henceforth \emph{Savage-Winkler} (see also 
Knuth\cite[p.~89]{knut:vfour-fascicle-two}).  It has
all $k$-of-$N$ codes appearing nearly together---interleaved
with codes of Hamming weight $k-1$ or $k+1$.  Consequently, 
successive $k$-of-$N$ codes have Hamming distance 2---just like the common/reflected Gray codes.

\begin{figure}
\begin{center}
  \subfigure[Run-length histogram]{\includegraphics[width=0.49\columnwidth]{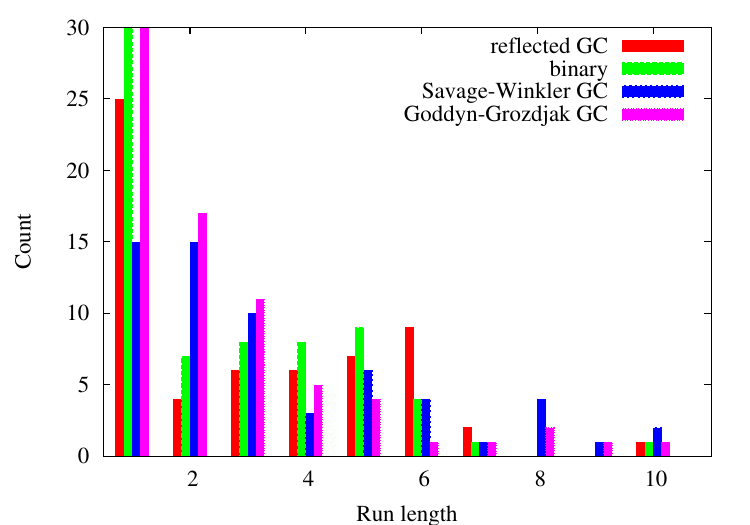}\label{fig:runlengths-kofN}}
  \subfigure[Probability within a long run, 2-of-8]{\includegraphics[width=0.49\columnwidth]{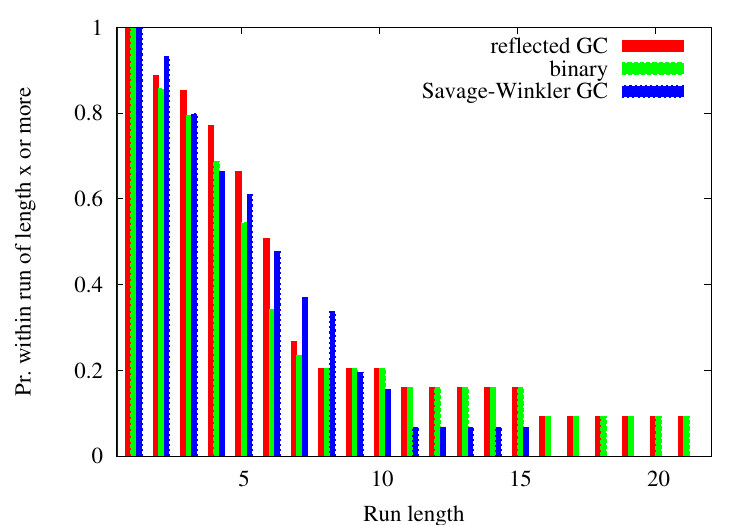}\label{fig:runlengths-cumulative-kofN}}
  \subfigure[Probability within a long run, 3-of-20]{\includegraphics[width=0.49\columnwidth]{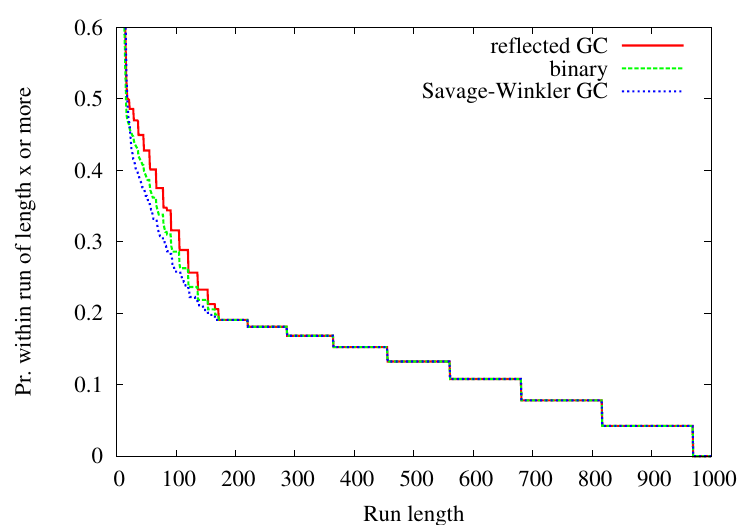}\label{fig:runlengths-cumulative-threeofN}}
  \subfigure[Probability within a long run, 4-of-14]{\includegraphics[width=0.49\columnwidth]{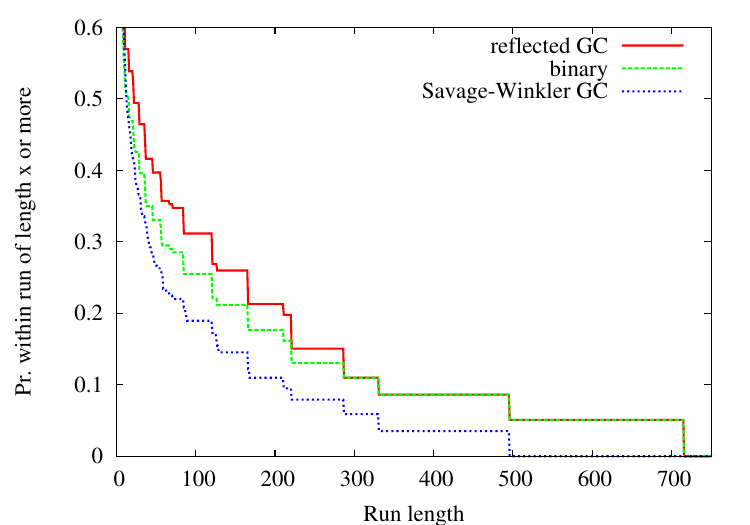}\label{fig:runlengths-cumulative-fourofN}}

\caption{\label{fig:runlengths-two-of-eight}
 Effect of various
orderings of the $k$-of-$N$ codes.
Top left: Number of runs of length $x$, 2-of-8 codes.  For legibility, we omit
counts above 30.  Goddyn-Grozdjak GC had 42 runs of length 1, binary had 32, and
random had 56.  Binary, reflected GC and Savage-Winkler had a run of length 15,
and reflected GC and binary had a run of length 22.
Remainder: Probability that a randomly-chosen bit from an uncompressed index
falls in a run of length $x$ or more. Goddyn-Grozdjak and the random
ordering were significantly worse and omitted for legibility.  
In~\ref{fig:runlengths-cumulative-threeofN}, when the techniques differed,
reflected GC was best, then binary, then Savage-Winkler GC. 
}
\end{center}
\end{figure}

The run-length distributions of the various codes are
heavily affected when we limit ourselves to $k$-of-$N$ codes.
This is illustrated by Fig.~\ref{fig:runlengths-two-of-eight}, where we
examine the run lengths of the 2-of-8 codewords, as ordered
by various Gray codes.  The code noted as Goddyn-Grozdjak
was obtained by inspecting a figure of 
Knuth~\cite[Fig.~14d]{knut:vfour-fascicle-two}; some discussion in
the exercises may indicate the code is due to Goddyn and 
Grozdjak~\cite{godd:long-run-codes}.
%

From Fig.~\ref{fig:runlengths-kofN}, we see that run-length distributions vary considerably
between codes.
(These numbers are for lists of $k$-of-$N$ codes without 
repetition; in an actual table, attribute values are repeated
and long runs are more frequent.) 
 Both Goddyn-Grozdjak GC and the random listing stand out as 
having many short runs. However, the important issue is whether the codes
support many \emph{sufficiently long} runs to get compression benefits. 

Suppose we list all 
 $k$-of-$N$ codes. 
Then, we randomly select a single
bit position (in one of the $N$~bitmaps). 
 Is there a good chance that this bit
position lies within a long run of identical bits? 
For 2-of-8, 3-of-20 and 4-of-14, we computed these probabilities (see Fig.~\ref{fig:runlengths-cumulative-kofN}, \ref{fig:runlengths-cumulative-threeofN}~and~\ref{fig:runlengths-cumulative-fourofN}).
 Random ordering and the Goddyn-Grozdjak GC ordering
were significantly worse and they have been
omitted. From these figure, we see that standard reflected Gray-code
ordering is usually best, but ordinary lexicographic ordering is often able to provide long runs.  
Thus, we might expect that binary allocation will lead to few dirty 
words when we index a table.

\begin{figure}
  
  \subfigure[1000 3-of-20 codes]{\includegraphics[width=0.49\columnwidth]{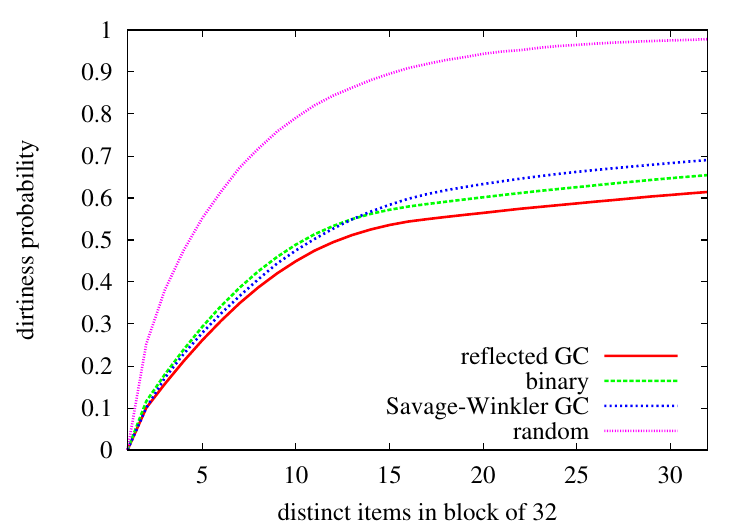}\label{fig:three-of-thirteen-locality}}
  \subfigure[1000 4-of-14 codes]{\includegraphics[width=0.49\columnwidth]{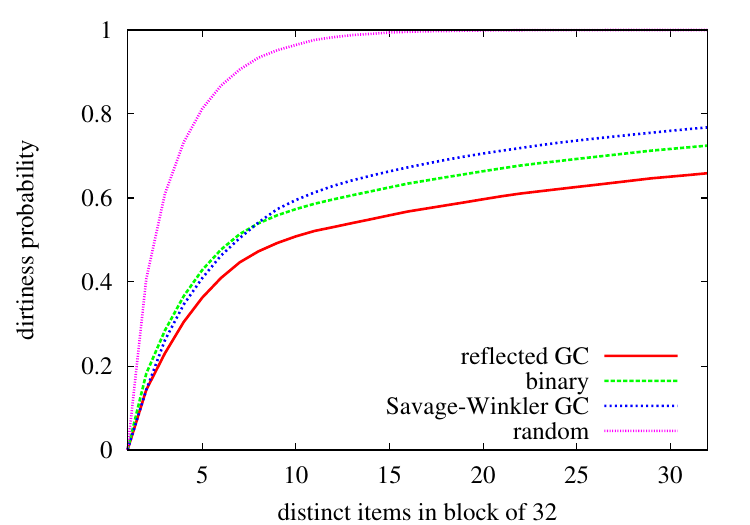}\label{fig:four-of-fourteen-locality}}
\caption{\label{fig:adjdirty-thousand}
Probabilities that a bitmap will contain a dirty word, when
several consecutive (how many: $x$-axis)
 of 1000 possible
distinct $k$-of-$N$ codes 
 are found in a 32-row chunk.
Effects are shown for values with $k$-of-$N$ codes that
are adjacent in  reflected GC, Savage-Winkler GC, binary or
random order.
}
\end{figure}

\paragraph{Minimizing the number of dirty words}
\killforgood{
Consider a bit 
in a $w$-bit word, prior to compression.
If that bit is not in a run of length $w$~identical bits or more, its word is dirty.
Even if it is in a run of length exactly $w$~bits, there is only a small
($1/w$) chance that the boundaries of the run coincide with the word
boundaries.  This chance rises for runs of length $w+1$, $w+2$, etc.
\suggestwecut{However, it is difficult to visualize the effect ---in terms of the
number of dirty words---of the various codes, even knowing the effects
on the run lengths.   We can assess the effect on dirty words by
experimental means, however.} \daniel{Owen had something here about
how it is difficult to visualize this.}
\owen{It was a lame attempt to fill a flow/concept gap that I think
still exists here.  We need some kind of segue.}
}
For a given column, suppose that in a block of 32~rows, we have $j$~distinct attribute values.
We computed the average number of bitmaps
whose word would be dirty (see Fig.~\ref{fig:adjdirty-thousand},
where we divide by the number of bitmaps).
Comparing $k$-of-$N$ codes that were adjacent in GC ordering against
$k$-of-$N$ codes that were lexicographically adjacent, the difference was
insignificant for $k=2$.   However, GC ordering is substantially better
for $k>2$, where bitmaps are denser. The difference between codes becomes
more apparent when many attribute values share the same word.  Savage-Winkler
does poorly, eventually being outperformed even by lexicographic 
ordering.
Selecting the codes randomly is disastrous. 
Hence, sorting part of a column---even one without long runs of identical
values---improves compression for $k>1$.






\subsection{Choosing the column order}\label{sec:dimOrder}

%
%
 
Lexicographic table sorting
uses the $i^\mathrm{th}$ column as the $i^{\mathrm{th}}$ sort key: it
uses the first column as the main key, the second column to break ties
when two rows have the same first component, and so on.
Some column orderings
may lead to smaller indexes than others. 

We model
the \emph{storage cost}  of a bitmap index
as
the sum of the number of dirty words and 
the number of sequences of identical clean words (1x11 or 0x00).
If a set of $L$~bitmaps has $x$~dirty words, then there are at most
$L+x$~sequences of clean words; the storage cost is at most $2x+L$.
This bound is tighter for sparser bitmaps.
Because the simple index of a column has at most $n$~1-bits, it has
at most $n$~dirty words, and thus, the storage cost is at most 
$3n$. The next proposition shows that the storage cost 
of a sorted column is bounded by $5n_i$.


\begin{proposition}\label{prop:sortingbenef}
Using GC-sorted consecutive $k$-of-$L$ codes, a sorted column with $n_i$ distinct values has no more than $2n_i$~dirty words,
and the storage cost  is no more than $4n_i+\lceil k n_i^{1/k}\rceil$.
\end{proposition}
\omitproof{%
\begin{pf}
Using $\lceil k n_i^{1/k}\rceil$~bitmaps is sufficient to represent $n_i$~values.
Because the column is sorted, we know that the Hamming distance of the bitmap
rows corresponding to two successive and different attribute values is 2. Thus
every transition creates at most two dirty words. There are
$n_i$ transitions, and thus at most $2n_i$ dirty words. This proves the result.
\end{pf}}

\begin{figure}\centering
\includegraphics[width=0.5\columnwidth]{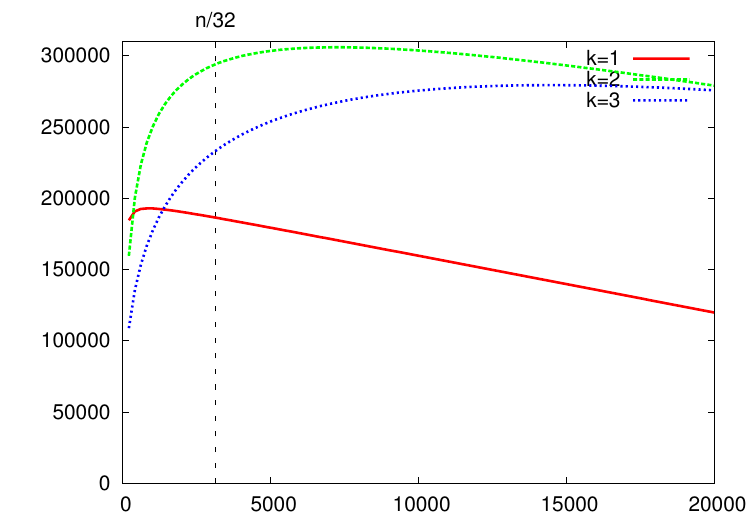}
\caption{\label{fig:theorygain}Storage gain 
in words for sorting a given column with $100,000$~rows and various number of attribute values
($2\delta(kn,\lceil k n_i^{1/k}\rceil,n)- 4n_i$ ).}
\end{figure}

For $k=1$, Proposition~\ref{prop:sortingbenef} is true irrespective of the order
of the values, as long as identical values appear sequentially.
Another extreme is to assume that all 1-bits are randomly 
distributed. Then sparse bitmap indexes
 have
$\approx \delta(r,L,n)=(1-(1-\frac{r}{Ln})^w)\frac{Ln}{w}$~dirty words where $r$ is the number of 1-bits, $L$ is
the number of bitmaps and $w$ is the word length ($w=32$).
Hence, we have an approximate storage cost of 
$2\delta(r,L,n)+\lceil k n_i^{1/k}\rceil$. 
The \emph{gain} of column $\mathcal{C}$ is the
difference between the expected storage cost of a randomly row-shuffled $\mathcal{C}$,
minus the storage cost of a sorted $\mathcal{C}$.
We estimate the gain  by
$2\delta(kn,\lceil k n_i^{1/k}\rceil,n)- 4n_i$ 
(see Fig.~\ref{fig:theorygain}) for columns with uniform histograms.
The gain is modal: it increases until a maximum is reached and then it
decreases.
The maximum gain is reached at $\approx \left( n(w-1)/2 \right)^{k/(k+1)}$:
%
for $n=100,000$ and $w=32$, the maximum is reached at $\approx 1\,200$ for $k=1$ and
at $\approx 13\,400$ for $k=2$.
Skewed histograms have a lesser gain for a fixed cardinality $n_i$. 

\begin{figure}
\centering
\subfigure[Uniform histograms with cardinalities 200, 400, 600, 800\label{fig:uni-order}]{%
\includegraphics[width=0.48\columnwidth]{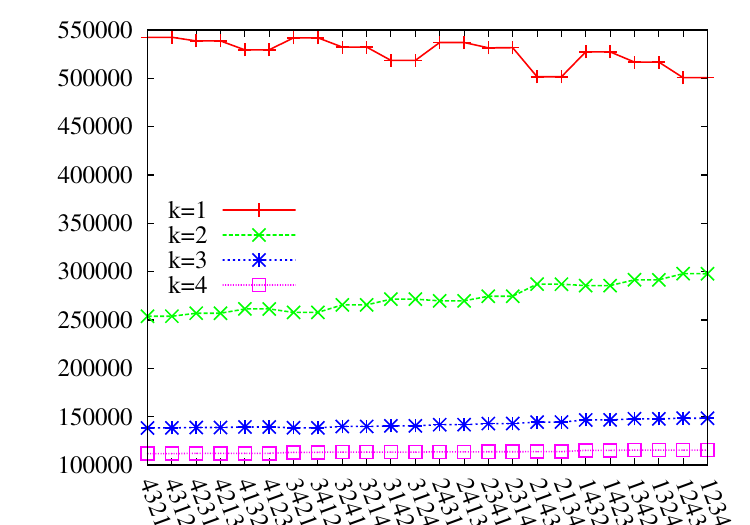}
}
\subfigure[Zipfian data with skew parameters 1.6, 1.2, 0.8 and 0.4\label{fig:zipfwide-order}]{%
\includegraphics[width=0.48\columnwidth]{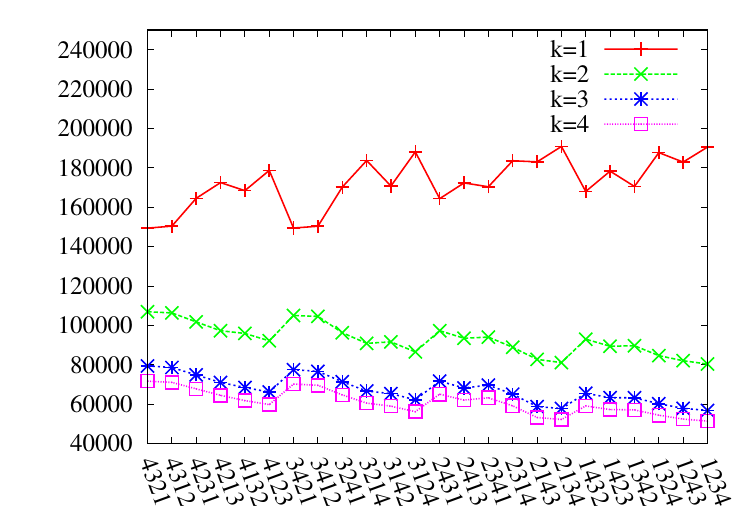}
}
\caption{Sum of EWAH bitmap sizes in words for various dimension orders on synthetic
 data ($100,000$ rows). Zipfian columns have 100 distinct values.\label{fig:toyorderings}
Ordering ``1234'' indicates ordering by descending skew (Zipfian) or
ascending cardinality (uniform).
%
}
\end{figure}

Lexicographic sort divides the $i^{\mathrm{th}}$~column 
into at most $n_1 n_2\cdots n_{i-1}$~sorted 
blocks.
 Hence, it has at most $2 n_1 \cdots n_i$~dirty
words.
When the distributions are skewed, the $i^{\mathrm{th}}$~column will have
blocks of different 
lengths and their ordering depends
 on how the columns are ordered.  
 	For example, if the first dimension
is skewed and the second uniform, the short blocks will be clustered,
whereas the reverse is true if columns are exchanged.
Clustering
the short blocks, and thus the dirty words, increases compressibility. 
Thus, it may be preferable to put skewed columns in the first positions
even though they have lesser sorting gain. 
To assess these effects, we generated data with 4~independent  columns:  using
uniformly distributed dimensions of different sizes (see Fig.~\ref{fig:uni-order}) and 
using same-size dimensions of different skew (see Fig.~\ref{fig:zipfwide-order}). 
We then determined
the Gray-Lex 
index size---as measured by the sum of bitmap sizes---for each of the 4! different dimension orderings.
Based on these results, for sparse indexes ($k=1$), dimensions should be ordered from least  to most skewed, and
from smallest to largest;
whereas the opposite is true for $k>1$. 

A sensible heuristic might be to sort columns by increasing density ($\approx n_i^{-1/k}$). 
However, a very sparse 
column  ($n_i^{1/k}\gg w$) 
will not benefit from sorting 
(see Fig.~\ref{fig:theorygain})
and should be put last. Hence, we use the following heuristic: columns are sorted in 
 decreasing order with respect to $\min(n_i^{-1/k}, (1-n_i^{-1/k})/(4w-1))$: this function 
 is maximum  at density $n_i^{-1/k}= 1/(4w)$ and it goes down to zero as the density goes to 1. 
 In Fig.~\ref{fig:uni-order}, this heuristic makes the best choice for all values of $k$.
We  consider this heuristic further
in \S~\ref{sec:column-exper}. 

\killforgood{
\wascut{%
If we only sort using the first column as a sorting key and have
 statistically independent columns,
a column ordering with maximal first-column gain has minimal expected storage cost.
But when using lexicographic sorting, 
we might want to choose a column with little sorting gain as
the first column.
If the first column has attribute values with frequencies $f_1^{(1)}, f_2^{(1)},\ldots$, then
the second column has at most $\sum_{z=1}^{n_1} \min(k f_z^{(1)}, 2 n_2)$~dirty words assuming GC-sorted
$k$-of-$L$ codes are used.}
\wascut{---for the $j^{\mathrm{th}}$~column, we should consider
the frequencies of the $j-1$~tuples made by the entries of the $j-1$ first columns.
To see why this result must hold, consider a sequence of $f_z^{(1)}$~rows
for which the value of the first column is constant. Because the
values of the second column are sorted and there are at most 
$n_2$ of them, we have at most $2n_2$ dirty words by Proposition~\ref{prop:sortingbenef}.
Also, there are at most $k f_z^{(1)}$~1-bits, and thus no more than 
$k f_z^{(1)}$~dirty words.} 
\wascut{\daniel{This whole thing is awfully vague:}
This analysis suggests the following
remarks:
\begin{itemize}
\item If the histogram sizes ($n_i$) are small compared to
the frequencies of the individual attribute values ($n_i \leq k f_z^{(i-1)}$), then we can
bound the storage cost for $k=1$ of two columns by 
$4n_1+4 n_1 n_2 +n_2+\lceil k n_1^{1/k}\rceil+\lceil k n_2^{1/k}\rceil$ and it 
is maybe best to put the column with the largest histogram first.
\cut{%
$4n_1+\lceil k n_1^{1/k}\rceil+
2 \sum_{z=1}^{n_1} \min(k f_z^{(1)}, 2 n_2) +\lceil k n_2^{1/k}\rceil=
4n_1+4 n_1 n_2 +n_2+\lceil k n_1^{1/k}\rceil+\lceil k n_2^{1/k}\rceil$. Hence, it is best to put the column with the largest
histogram first. This is consistent with maximizing the gain of the first column.}
\item If  columns have relatively infrequent items ($f_z^{(i-1)}\leq n_i/k $),
then we expect that only the first column will benefit from the sorting, 
and it is best to maximize its gain.
\item However, when the first dimension has infrequent attribute
values ($f_z^{(i-1)}\leq n_i/k $) whereas the second dimension
has frequent attribute values ($f_z^{(i)}\leq n_{i-1}/k $), it might be
best to put the second dimension first. 
\end{itemize}
\daniel{End of evil vagueness?}}
\cut{%
\begin{figure}\centering
\includegraphics[width=0.8\columnwidth]{log2dskewedforowenk2}
\daniel{\\For $k=1$, it does not work:\\
\includegraphics[width=0.8\columnwidth]{log2dskewedforowen}
}\\
\daniel{A similar experiment without skew, but with different cardinalities does not work for any $k$:\\
\includegraphics[width=0.8\columnwidth]{log2duniforowenk2}
}
\caption{\label{fig:skewedforowen}Storage gain 
in words for sorting skewed tables with $n=10000$ and different skews ($k=2$)}
\end{figure}
}
\cut{%
\subsubsection{Uniform histograms}
\label{subsubsectionunihisto}
\daniel{Daniel has the conjecture that for uniform histograms, 
sorting by the gain is good enough. This is \textbf{proven} by the
following figure:\\
\includegraphics[width=0.8\columnwidth]{uniformhistograms}\\
In other words, if you can assume that the histograms are
flat, just sort the columns in inverse order of the gain and
you will be fine. We should be able to prove this analytically.}
}
\cut{%
\subsubsection{Simple bitmap indexes}
\daniel{Daniel had the conjecture that for $k=1$, basically, you do not need
to worry about the effect of the bias. I proved this false with
the simpleindex.sh script. So we can drop this subsubsection.}
}
\cut{%
\subsubsection{Measuring the bias of the histogram}
\daniel{A natural bias measure is the difference between the
gain measured above from $k$ and $n_i$, with a uniform histogram assumption,
and the (lower) gain measured by taking into account the full histogram.}
\daniel{Indeed, using the result from the subsubsection~\ref{subsubsectionunihisto},
we have that when the bias is zero, according to this measure,
then you should just sort by gain because the distributions are uniform.}
\daniel{A wild conjecture is that whatever a column loses in gain
due to its bias, it makes it up by allowing the following columns
to be better sorted.}
}
\cut{\owen{Probably obsolete discussion: Looking at experimental(skewed), it appears that the conclusion
for $k=1$ is unaffected: n/32 gets best gain.  For k=2 uniform, and k=3
(skewed or uniform),  your theoretical curves suggest bigger is better.
But for k=2 skewed, there appears to be a gentle peak around 1500.\\
Daniel's figures 3b can be compared to Owen's~\ref{fig:zipfs2-order}
and 3c to~\ref{fig:zipfs10}.  For k=1, Owen's conclusion can be that
the same orders are suitable but the difference is less extreme
when skew is high.  Daniel's does not seem to contradict this: gain
has similar peak location, but is less extreme.  For both Daniel and
Owen, ramping up the skew does not change the conclusion for $k>1$
that big dimensions should go first.  For Owen, k=2, the difference
between the best and worst ordering was a factor of 2, regardless of
large or small skew.  Daniel's result would have lead to the guess that
ordering was less important for higher skew.  Hmm.}
}
\cut{\owen{Experimentally, skew seems to matter a lot (see pix toward end), but
 affects $k=1$ in an opposite way from how it affects $k>1$.
 For sanity, maybe we are stuck with a distribution oblivious model here,
 but somehow, eventually, we ought to try to knit the skew effects 
(whether we can  explain it mathematically or just describe what we see)
together with these results relating table length, $k$ and dimension
size.}}
\cut{\daniel{I have added skewed distributions in my figures. It seems that column
order is less important if the distributions are very skewed. Which makes sense
if you think about it. Imagine a distribution where all values occur once, except
for one that is repeated often. If the number of distinct attribute values,
this is nearly as having just one attribute value, and then, clearly, column order
does not matter.}}
%
%
%
%
\cut{\daniel{An obvious question maybe is why not sort by gain as a heuristic? This would tend
to optimize mostly the first few dimensions. But looking at table 7 from BDA,
it looks like sorting by gain might do the right thing. 
\owen{After sorting by some first dimension, you essentially create a
bunch of sort-able sub problems, one for each attribute value in the 
first dimension.  To choose the second sort dimension, you should not
apply the original gain formula.  Instead, the number of rows should be
adjusted to be suitable for the sizes of the sub-problems.  We could 
choose average, which is appropriate if the first dimension was uniform.
However, if the first dimension was fairly big and uniform, it does not
much matter how you order the remaining dimensions.  So try to help
skewed ones.  I suggest that you let $n$ be the number of occurrences
of the most frequent value in dimension 1, for choosing dimension 2.
And for choosing dim 3, let $n$ be the number of occurrences of the 
most frequent pair
$(v1,v2)$ where $v1$ is from the first dimension, $v2$ is from the
second dimension. Etc.\\
If you just generate gains for all dims using the same $n$, then 
the choice made for the subproblems will not avoid excessively large
dimensions.
}
The histogram-aware
formula is really not that hard:
$\sum_{j=1}^{\lceil k n_i^{1/k}\rceil} 2 (1-(1-f_j/n)^w) n/w+\lceil k n_i^{1/k}\rceil-
2n_i$ (assuming I did not screw this up) where $f_j$ represents the per-bitmap
histogram.
One can validate that the formula is roughly sensible by taking each
column, shuffling it, computing the size, the subtracting the 
size of the index over the sorted column.
}
}
\cut{%
In Section~\ref{sec:exper-ordering}, we evaluate the following heuristics:
\daniel{For each heuristic, we could plot a corresponding cost measure that it
seeks to minimize, something like $n^{d-1} \min(n_1, n/32)+ n^{d-2} \min(n_2, n/32)+\ldots$.  }
\begin{enumerate}
\item sort the columns in decreasing order with respect to $\min(n_i,k^2 n (w-1) / (2\max_{j=1,\ldots,d} n_j)$;
\item sort the clumsy in decreasing order with respect to $\min(n_i, n/32)$ \daniel{If we are going
to set an arbitrary point such as $n/32$, why not pick $\approx \left( \frac{w-1}{w} \right)^{\frac{k}{k+1}} n^{k/(k+1)}$?};
\item sort the clumsy in decreasing order with respect to the number of attribute values
occurring more than 32~times.
\end{enumerate} 
\cut{As a heuristic, in Section~\ref{sec:exper-ordering} we evaluate the following approach, which 
requires knowing only the schema information and the number of rows, $n$:
choose as the primary sort key the dimension $D_i$ with largest
$n_i$, as long as $n / n_i \geq 32$.  
\owen{vaguely resembles the histo approach sketched at the end of 4.3 of BDA, I guess}
The second sort key is the 
next largest dimension, and so forth.  This
is our \textbf{ordering heuristic 2}.  Another and similar 
heuristic~\cite{bda08} requires
computing histograms: we favour the dimensions in order of
descending size, where we consider only values that have more than 32 occurrences. 
 This is \textbf{ordering heuristic 3}.}
}
\cut{%
If $k=1$, favoured dimensions should
have low skew (calculate with statistical methods),
except that it should not begin with a small dimension followed
by a dimension that does not have some highly frequent
members.    
If $k>1$, favoured dimensions should have high skew and large size.
\owen{Some combination is required?}
As $k$ increases, size becomes less significant.
\owen{Some combination of skew and the kth root of the dim size,
representing the number of bitmaps?}
This leads to \textbf{ordering heuristic 4:}  For $k=1$, order
dimensions by increasing skew (need a quick test for that).
For $k>1$, for each dimension $D$ we
compute $g_D |D|^{1/k}$, where $g_D$ is $<$some statistical test that
estimates skew in a reasonable way$>$.  Order dimension by
decreasing value of this product.
}
} 

\subsection{Avoiding column order}  

Canahuate et al.~\cite{pinarunpublished} propose to permute bitmaps individually prior to sorting,
instead of permuting table columns. We compare these two strategies
experimentally in \S~\ref{sec:bitmap-reordering-exper}.

\killforgood{
Bitmap ``interleaving'' schemes are possible: if dimension $i$ has bitmaps $b_{i,1}, b_{i,2}, \ldots b_{i,L_i}$, then we could  sort the index lexicographically
by \begin{eqnarray*}b_{1,1}, b_{2,1}, \ldots, b_{D,1}, b_{1,2}, b_{2,2}, \ldots\end{eqnarray*}
}

\killforgood{
%
%

Although we suggest first sorting the  table and then encoding
the sorted table into a compressed bitmap index, our approach appears
less flexible than first materializing an uncompressed index,
then sorting, then compressing. 
However, the B-tree method sketched in \S~\ref{sec:sorting-rows}
yields an equivalent result, enabling a more equal treatment of
(primary sort key, second sort key, and so forth).  
\cut{Except for the first
few dimensions, many dimensions do not have a large effect on 
the sort.   We can treat dimensions more equally if we consider the
individual bitmaps arising from the dimensions.}
Suppose dimension $i$ leads to bitmaps $b_{i,1}, b_{i,2}, \ldots b_{i,L_i}$.
One ``interleaving'' scheme for would be to sort the index lexicographically
by $b_{1,1}, b_{2,1}, \ldots, b_{D,1}, b_{1,2}, b_{2,2}, \ldots$.
Canahuate et al.\  similarly propose
to order individual bitmaps by their densities~\cite{pinarunpublished}. 
}

\killforgood{ 

\owen{I would like to construct a scenario where bitmap-wise
column reordering must be much better than column wise.  Or find
a reason why this cannot be done. My idea, which I have not made
work, is to suppose some specific
individual attribute values are highly correlated between different
dimensions, but the dimensions as a whole are not highly correlated
(just what they get from these specific attribute values).
So far, whenever I seem to have an idea, I can always suppose an
appropriate ordering on the attribute values and obtain a good result
without per-bitmap reordering. Maybe this can be formalized to show
something....I dunno.}

\owen{There was something on QDBM here, but the discussion earlier
(when the sparse comparator was given) seems to already have all that
is needed.  I have commented it out.}
}
\killforgood{
\notessential{%
It \emph{is} possible to sort the  table in a bitmap-interleaved
fashion, without materializing the bitmap index first.   Any
general-purpose external-memory sorting algorithm may be used, providing
we may specify how rows are compared.   To compare rows $r_1$
and $r_2$, we temporarily convert the rows to their corresponding
lines in the bitmap index, and then compare the two lines as desired
(by Gray code ordering, or lexicographically according to the densest
bitmaps, etc.)  Although we know that O($n \log n$) comparisons will be
sufficient, each comparison can be expensive:  it entails encoding
into a long (probably sparse) vector.  \owen{we may have
already hit this enough earlier} A similar idea is to store
each row and (using sparse vector methods) its bitmap line, in an
external memory data structure for ordered sets: we used the B-trees 
provided by QDBM~\cite{qdbm}.
\owen{we could translate the stuff at end of BDA section 3 to elaborate.}
The key-comparison routine is then based on the stored bitmap line,
which only needs to be calculated once.  The disadvantage is that the
external memory data structure needs storage comparable to the
 table and a (row-) compressed bitmap index.  This implementation
was initially used for our GC sorting, but it proved at
least 100 times slower than lexicographically sorting.
}
}


As a practical alternative to lexicographic sort and column
 (or bitmap) reordering, we introduce \emph{Frequent-Component} (FC) sorting, which
uses histograms to help sort
without bias from a fixed dimension ordering.
In sorting, we compare the frequency of the $i^\mathrm{th}$ most frequent attribute values in 
each of two  rows without regard (except for possible tie-breaking)
to which columns they come from. For example, consider the following
table:
\begin{center}
\begin{tabular}{ll}
cat & blue \\
cat &red \\
dog &green\\
cat &green \\
\end{tabular}
\end{center}
We have the following  (frequency, value) pairs: (1,blue), (1,red), (1,dog), (2,green), and (3,cat). 
For two rows $r_1$ and $r_2$, $<_{\mathrm{FC}}$ first compares
 $(f(a_1),a_1)$ with $(f(a_2),a_2)$, where $a_1$ is the least
frequent  component in $r_1$ and $a_2$ is the least frequent 
component in $r_2$---$f(a_i)$ is the frequency of the component $a_i$.  Values $a_1$ and $a_2$ can be from different columns.  Ties are broken by the 
second-least-frequent components in $r_1$ and $r_2$, 
and so forth. Hence, the sorted table in our example is
\begin{center}
\begin{tabular}{ll}
dog & green \\
cat & blue  \\
cat & red  \\
cat & green.
\end{tabular}
\end{center}

With appropriate pre- and post-processing, it is possible to implement FC using a standard sorting utility such as Unix \textbf{sort}.
 First, we sort the components of each row of the  table
into ascending frequency.  In this process, each component is
replaced by three consecutive components, $f(a)$, $a$, and $\textrm{pos}(a)$.
The third component records the column where $a$ was originally found.
In our example, the table becomes 
\begin{center}
\begin{tabular}{ll}
(1,dog,1) & (2,green,2)\\
(1,blue,2) & (3,cat,1) \\
(1,red,2)  & (3,cat,1)\\
(2,green,2) & (3,cat,1).
\end{tabular}
\end{center}
Lexicographic sorting (via \textbf{sort} or a similar utility) of rows
follows, after which each row is put back to its original value
(by removing $f(a)$ and storing $a$ as component $\textrm{pos}(a)$).
\section{Picking the right $k$-of-$N$}\label{sec:multi}

Choosing $k$ and $N$ are important
decisions.  We choose a single $k$ value for all dimensions\footnote{%
Except that for columns with small $n_i$, we automatically adjust
$k$ downward when it exceeds the limits noted at the end of 
\S~\ref{sec:bitmapIndexes}.}, leaving
the possibility of varying $k$ by dimension as future work.
Larger
values of $k$ typically lead to a smaller index and a faster construction
time---although we have 
observed cases where $k=2$ makes a larger index. 
However, query times increase
with $k$: there is a construction time/speed
tradeoff.

\subsection{Larger $k$ makes queries slower}\label{sec:slowquery-theory}
We can bound the additional cost of queries. Write ${L_i \choose k}=n_i$.
A given $k$-of-$L_i$ bitmap is the result of an OR operation over at most 
$k n_i/L_i$~unary  
bitmaps by the following proposition.

\begin{proposition}
In $k$-of-$N$ encoding, each attribute value is linked to 
$k$~bitmaps, and each bitmap is linked to at most $\frac{k}{N} {N \choose k}$~attribute values.
\end{proposition}
\begin{pf}
There are ${N \choose k}$~attribute values. Each attribute
value is included in $k$~bitmaps. The bipartite graph 
from attribute values to bitmaps has $k {N \choose k}$~edges.
There are $N$~bitmaps, hence $\frac{k}{N} {N \choose k}$~edges per bitmap.
This concludes the proof.
\end{pf}

Moreover, $n_i = {L_i \choose k} \leq  (e \cdot L_i/k)^k$ by a standard inequality, so that $L_i/k \geq n_i^{1/k} / e$ or  $k/L_i \leq  e \cdot  n_i^{-1/k} < 3 n_i^{-1/k}$. Hence, $kn_i/L_i < 3 n_i^{(k-1)/k}$.

 Because 
$|\bigvee_i B_i|\leq \sum_i | B_i|$, 
the expected size of such a $k$-of-$L_i$~bitmap 
is no larger than  $3 n_i^{(k-1)/k}$~times the expected size of a unary bitmap. 
%
%
%
%
A query looking for one attribute value will have to AND together
$k$ of these denser bitmaps. 
The entire ANDing operation can be done (see the end of \S~\ref{sec:compression})
by $k-1$ pairwise ANDs that produce intermediate results whose EWAH sizes are
increasingly small:
$2k-1$~bitmaps are thus processed. 
Hence, the expected time complexity of  an equality query on a dimension of size
$n_i$ is no more than  $3(2k-1) n_i^{\frac{k-1}{k}}$~times higher than the expected
cost of the same query on a $k=1$ index.
(For $k$ large, we may use see Algorithm~\ref{algo:genrunlengthmultiplefaster} to substitute 
$\log k$ 
for the $2k-1$ factor.)

For a less pessimistic estimate of this dependence, observe that indexes
seldom increase in size when $k$ grows.  We may conservatively assume that
 index size is unchanged when $k$ changes.
Therefore, the expected size of one bitmap grows as the reciprocal of
\owen{or something about inversely proportional? Was ``as one over''
which I found hard to parse here.}
the number of bitmaps ($\approx n_i^{-1/k}/k$), 
leading to queries whose cost is proportional to 
$\approx (2k-1)  n_i^{-1/k}/k = (2-1/k)n_i^{-1/k}$.  Relative to the cost for $k=1$, which
is proportional to $1/n_i$, we can say that 
increasing $k$ leads to
queries that are 
                 $(2-1/k)n_i^{(k-1)/k}$ 
times more expensive than
on a simple bitmap index.

For example, suppose $n_i = 100$.  Then 
going from
$k=1$ to $k=2$ should increase query cost about 15 fold 
but no more than 90 fold. 
%
In summary, the move from $k=1$ to anything larger can have a 
dramatic negative effect on query speeds.  Once we are at $k=2$,
the incremental cost of going to $k=3$ or $k=4$ is low:
whereas the ratio $k=2 : k=1$ goes as $\sqrt{n_i}$, the ratio
$k=3 : k=2$ goes as $n_i^{1/6}$.
We investigate this issue experimentally in \S~\ref{sec:queries-exper}.

\killforgood{
\cut{%
We would, at least, have to read all these
bitmaps, and so we might expect equality query times to grow with $k^2$.  
However, the total size of the index may be smaller.
To understand this tradeoff, we focus on primarily on how $k$ affects 
the index size.  
\textbf{daniel analysis to replace above.}
{Let me redo the analysis from scratch. Make it clear
that this is for equality queries only.
Ok. We know that taking AND between two bitmaps can be done
in time proportional to the sum of the compressed sizes of the bitmaps.
For the AND operation between $k$ bitmaps, a pessimistic bound
tells you that the operation is in time proportional to $(2k-1)B$
where $B$ is the size of the bitmaps (here, we should state
precisely the algorithm we use: merge two compressed bitmaps into 
another compressed bitmaps and so on). This is pessimistic because
AND operations between bitmaps tends to generate smaller and smaller bitmaps.
 Ok, we have $k n_i^{1/k}$ bitmaps where $n_i$ is the number of
attributes. We know
that the total size of the bitmap index hardly ever increases with $k$ and
usually diminishes. Let us be pessimistic and assume that the size
remains the same as $k$ increases. Then the size of each bitmap
goes as $\frac{S}{k n_i^{1/k}}$. Hence, my pessimistic bound
puts the query time proportional to 
$\frac{(2k-1) S}{k n_i^{1/k}}$. So, for big dimensions, there might
be a huge difference between the various $k$, but less so for the smaller
dimensions. Of course, the size of the index $S$ diminishes with $k$
and the $(2k-1)B$ is likely to be quite pessimistic for $k=4$.
Suppose we are dealing with a small dimension $n_i=100$, the
the ratio of the query speed of $k=2$ to $k=1$ is 15
whereas $k=3/k=1$ is  35 and $k=4/k=1$ is 55. Increasing the dimension
size considerably, say $n_i=10000$, we get more drastic differences:
ratio $k=2/k=1$ is 150, $k=3/k=1$ is 774 and $k=4/k=1$ is 1750.
}}}
\killforgood{
To test it out I ran the following experiment. I generated a 1d table
with a uniformly distributed histogram. I set the table length to 100\,000
rows and I varied the number of distinct values and $k$. I then
plotted the time (in seconds) to execute 1\,000 queries on one bitmap (alternatively,
you could say that the time reported is in millisec):\\
\includegraphics[width=0.8\columnwidth]{unisynthtimeversuscardi}\\
Then I plotted the $\frac{(2k-1) S}{nbrbitmaps}$ estimate (in words):\\
\includegraphics[width=0.8\columnwidth]{unisynthbitmapsizeversuscardi}\\
There is no agreement between the bound and the data,  but it is a pessimistic bound.
\daniel{Owen properly reminds us that hardware and OS-level caching will help
 the
looped queries, thus making our results a bit unrealistic. This should
be pointed out in the text. }
\daniel{I could easily plot something like $\frac{k S}{nbrbitmaps}$ which
would give us the amount of data being loaded up. It might be false to
equate it with the query time, but it certainly contribute to the
query time.}
}

\killforgood{
\cut{%
\owen{This is more the subject of one of the other manuscripts\ldots}
\owen{following paragraph is probably removable junk; DL has earlier put
some similar accounting.}
With $k=1$, there is a wide range on the number of dirty words:
Since the index will have $rc$~1-bits, it can have at most $rc$
dirty words. (Ignoring really long runs of zeros).  In the first
dimension.  The EWAH cost can be bounded by observing that
the worst case would have a run of zeros between every dirty word
in a bitmap,
as well as before the first and last dirty word.  For $k=1$,
a lower bound is that each of the bits must be accounted for.
Ideally, each bitmap would consist of a run of zeros,
a run of ones, and a final run of zeros.  It would require
3 words in  EWAH\@.  More realistically, we
would not have the run of ones aligned on word boundaries, and
thus a dirty word would would follow and precede the run of ones,
leading to
5 words per bitmap.
Thus, for $k=1$, our index EWAH size is  between $3N$ and 
$\min(3N, N+ 2*rc)$.
Repeating for $k=2$, we have a total of $2rc$ bits, and an
index EWAH size between $3 \Theta(\sqrt{N})$ (fixme) and 
$\min(3\Theta(\sqrt{n}), \Theta(\sqrt{N} + 4*rc)$
Thus, we see that as $k$ increases our bounds become broader.
\owen{It is easy to construct inputs meeting the lower bound, but
what about the upper bound?}
}
}


\subsection{When does a larger $k$ make the index smaller?}
Consider the effect of a length 100 run of values $v_1$, followed
by 100 repetitions of $v_2$, then 100 of $v_3$, etc.  Regardless of
$k$, whenever we switch from $v_1$ to $v_{i+1}$ at least two bitmaps
will have to make transitions between 0 and 1. 
Thus, unless the
transition appears at a word boundary,
 we create at least 2
dirty words whenever an attribute changes from row to row.  The
best case, where \emph{only} 2 dirty words are created, is achieved when $k=1$
for \emph{any} assignment of bitmap codes to attribute values.
For $k>1$ and $N$ as small as possible, it may be impossible to
achieve so few dirty words, or it may require a particular assignment
of bitmap codes to values.


Encodings with $k>1$ find their use when many (e.g. 15) attribute
values fall within a word-length boundary.  In that case, a
$k=1$ index will have at least 15 bitmaps with transitions
(and we can anticipate 15 dirty words).  However, if there were only
45 possible values in the dimension, 10 bitmaps would suffice
with $k=2$.  Hence, there would be at most 10 dirty words and
maybe less if we have sorted the data (see Fig.~\ref{fig:adjdirty-thousand}).  
\killforgood{In
fact, with 45 values in a dimension encoded with 2-of-10 encoding,
a random selection of 15 distinct codes has an expected number of 0.35
bits in common (see Proposition~\ref{numofexpectedbits}).}

\killforgood{%
removing attempt to quantify (not very helpful, not very right)

\owen{this has a lot in common with things DL added on June 9,
so there is room to save space here.}

\begin{proposition}\label{numofexpectedbits}
Given $x$ distinct values with $k$-of-$N$ encodings, the expected number
of bits where all codes agree is $formula(x,k,N)$.
Something like $N ( (1-k/N)^x + (k/N)^x)$ but not exactly.
\end{proposition}

\begin{pf}
Depends on formula.  Sanity: 
formula should be $0$ when $x \geq {N-1 \choose k}$.
Should be $k$ when $x=1$.

The probability that any given bit position is clean is the
sum of the probability that it is a clean 1 and the probability
that it is a clean 0, since these possibilities exclude one
another.

In a code, suppose that each bit position has an independent
probability of being 1, and it is $k/N$.  (This is obviously wrong,
since if $k$ is 2 and you know two other bit positions are 1, then
the probability of any remaining bit position storing 1 is 0.  Oh
well, this is a starting point.  Also, I want $x$ distinct codes
(no repetition)).   The probability that in $x$
random codes, there are only zeros [in a pre-specified bit position]
 is $(1-k/N)^x$, and the probability
there are only ones is $(k/N)^x$.
\end{pf}

\subsection{Comment} For our case study, evaluating
$N ( (1-k/N)^x + (k/N)^x)$ with $N=10,\ k=2,\ x=15$ we have
0.35; in other words, we expect 9.65 dirty words rather than 15.
This 0.35 is not very significant.  For a fixed small value of
$k$, this wrong formula is significant wrt N when N is large and
x is fairly small.  Eg, x=4, N = 20, k=2 we have about $.9^4 20$
or about 13.  13 clean words, hence 7 dirty words.   However,
simple coding would have only 4 dirty words.
}
%

\subsection{Choosing $N$}


It seems intuitive, having chosen $k$,
to choose $N$ to be as small as possible.   Yet, we have observed
cases where the resulting 2-of-$N$ indexes are much bigger
than 1-of-$N$ indexes.  Theoretically, this could be avoided 
if we allowed larger $N$, because one could aways append an additional
1 to every attribute's 1-of-$N$ code.  Since this would create one
more (clean) bitmap than the 1-of-$N$ index has, this
2-of-$N$ index would never be much larger than the 1-of-$N$ index.
So, if $N$ is unconstrained, we can see that there is never a
significant space advantage to choosing $k$ small.

Nevertheless,  the main advantage of $k>1$ is fewer bitmaps. 
We choose $N$ as small as possible.

\killforgood{%
\owen{you can obtain a k-of-N code from any 1-of-N by prepending k-1
columns of constant 1s.  Thus, if you do not constrain N, a k-of-N
code can always be almost as good (additive Theta(c*k) amount) as
good as a 1-of-N.  Question: if I constrain N to be as small as possible
(what we have been doing), can we make ourselves get a worse answer
for 2-of-N than 1-of-N?  I think so, from experiments.  Make a simple
manual example.}

\daniel{We have a funny k-of-N implementation where we pick a different
k per dimension when dimensions have too few attribute values. Daniel was
motivated by the following idea: if you have the same number of bitmaps, you
are better off with the lesser value of $k$.}\daniel{We could prepare a finer
analysis for DOLAP.}  this is in section 2

\owen{for
3-of-N coding and 4-of-N coding,  and -a weightgray, you obtain a smaller
index using  -K than -k on tweed6,3,4,52.}
\daniel{not necessarily faster performances.}
\owen{I am not sure if this is just a quirk of the heuristicness of
everything else that goes on.  In some cases, I think if you exchange the role of 1 and 0, what happens with -K and a ``too big'' value for the
dim size may be the (negation) of what happens with -k  and N-''too big''.
Since 1s and 0s are handled symmetrically in EWAH, this is not very
interesting.\\
We could look into this more, but I am not optimistic about this any more.
}
}

\section{Experimental results}\label{sec:Experiment}

We present experiments to assess the effects of various
factors (choices of $k$, sorting approaches, dimension
orderings) in terms of EWAH index sizes.   These factors
also affect index creation
and query times. We report real wall-clock times.

\subsection{Platform} 
Our test programs\footnote{\url{http://code.google.com/p/lemurbitmapindex/}.}
were written in C++ and compiled by GNU
GCC~4.0.2 on an Apple Mac Pro with two double-core Intel
Xeon processors (2.66\,GHz) and 2\,GiB of RAM\@.
Experiments used a 500\,GB SATA Hitachi disk (model 
HDP725050GLA360~\cite{hita:patiencedisk,dell:deskstar-user-guide}),
with average seek time (to read) of 14\,ms ,  
average rotational latency of 4.2\,ms,
and capability for sustained transfers at 300\,MB/s. 
This disk also has an on-board cache size of 16\,MB, and is formatted for
the Mac OS Extended filesystem (journaled).
Unless otherwise stated, we use 32-bit binaries.
Lexicographic sorts of flat files were done using GNU coreutils 
\textbf{sort} version 6.9.
For constructing all indexes, 
we used Algorithm~\ref{algo:owengenbitmap}
because without it, the 
index creation times were 20--100 times larger, depending on
the data set.


\subsection{Data sets used}\label{sec:datasets}



We primarily used four data sets, whose details are
summarized in Table~\ref{tab:caractDataSet}: Census-Income~\cite{KDDRepository}, DBGEN~\cite{DBGEN},
KJV-4grams, and Netflix~\cite{netflixprize}.
 DBGEN is a synthetic data set, whereas 
 KJV-4grams is a large list (including duplicates) of
4-tuples of words obtained from the
verses in the King James Bible~\cite{Gutenberg}, 
after stemming
with the Porter algorithm~\cite{275705} and removal of
stemmed words with three or fewer letters.
Occurrence of row $w_1, w_2, w_3, w_4$  indicates
that the first paragraph of a verse contains words $w_1$ through $w_4$, in this
order. 
KJV-4grams is motivated by research on 
Data Warehousing applied to 
                  text analysis~\cite{KaserKeithLemire2006}.
\killforgood{ 
This data is a scaled-up version of word co-occurrence cubes
used to study analogies in natural
language~\cite{TurneyML,KaserKeithLemire2006}.
}%
Each of column of KJV-4grams contains roughly 8~thousand distinct stemmed words.
The Netflix  table has 
4~dimensions:  UserID, MovieID, Date and Rating,
having cardinalities 
480\,189, 17\,770, 2\,182, and 5.
Since the data was originally supplied in 17\,700~small files
(one file per film), we concatenated them into a flat file
with an additional column for the film and randomized the
order of its rows using Unix commands such as
\verb+ cat -n file.csv | sort --random-sort | cut -f 2-+.
All files were initially randomly shuffled.

\begin{table}
     \caption{Characteristics of data sets used.
    }\label{tab:caractDataSet}
    \centering
    \begin{tabular}{l|rrrr|} \cline{2-5}
     & rows & cols & \rule{0mm}{1.1em} $\sum_i n_i$ & size \\ \hline
    \multicolumn{1}{|l|}{\textbf{Census-Income}} & 199\,523     & 42 & 103\,419   & 99.1\,MB   \\ 
    \multicolumn{1}{|r|} {4-d projection}        & 199\,523     & 4 & 102\,609   & 2.96\,MB   \\
    \multicolumn{1}{|l|}{\textbf{DBGEN}}         & 13\,977\,980  & 16 & 4\,411\,936   & 1.5\,GB \\ 
    \multicolumn{1}{|r|} {4-d projection}        & 13\,977\,980  & 4 & 402\,544   & 297\,MB  \\
    \multicolumn{1}{|l|}{\textbf{Netflix}}       & 100\,480\,507 & 4 & 500\,146   & 2.61\,GB \\
    \multicolumn{1}{|l|}{\textbf{KJV-4grams}}    & 877\,020\,839 & 4 &  33\,553   & 21.6\,GB  \\ 
\hline
    \end{tabular}

\end{table}

For some of our tests, we chose four dimensions with a wide range
of sizes.
For Census-Income, we chose \textit{age} ($d_1$), 
\textit{wage per hour} ($d_2$), \textit{dividends from stocks} ($d_3$)
and a numerical value\footnote{%
The associated metadata says this column should be a
10-valued migration code.} found in the $25^{\mathrm{th}}$ position ($d_4$). Their respective cardinalities 
were 91, 1\,240, 1\,478 and 99\,800.
For DBGEN, we selected dimensions of cardinality 7, 11, 2\,526 and 400\,000.
Dimensions are numbered by increasing
size: column 1 has fewest distinct values. 

\killforgood{%
\daniel{I am worried that TWEED might be a bad data set because I found
that the gain behaved similarly for all $k$'s when $n$ is small.
We see a big difference between $k=1$ and the $k>1$ for large values of $n$.
See Fig.~\ref{fig:theorygain} to see how $k=1$ differs quite a bit from the
others.}
We also used the small terrorism
data set, TWEED~\cite{tweed}, which has approximately 11~thousand rows. 
We projected
columns  
52 (Target), 
4 (Country),
3 (Year) and
6 (Acting Group), whose cardinalities were
11, 16, 53 and 297, respectively.
\daniel{Any reason to present the dimensions
in reverse cardinality order?}\owen{good catch.}
}

\subsection{Overview of experiments}
\label{subsection:efficiencyofEWAH}


Using our test environment,  our experiments assessed
\begin{itemize}
\item whether a partial (block-wise) sort could save enough time to
   justify lower quality indexes (\S~\ref{subsection:crazyblocks});
\item the effect that sorting has on index construction time (\S~\ref{subsection:IndexConstructionTime})
\item the merits of various code assignments
 (\S~\ref{sec:exper-sorting});
\item whether column ordering (as discussed in \S~\ref{sec:dimOrder})
  has a significant effect on index size (\S~\ref{sec:column-exper});
\item whether the index size grows linearly as the data set grows (\S~\ref{sec:exper-lexsorting});
\item whether bitmap reordering is preferable to our column reordering (\S~\ref{sec:bitmap-reordering-exper});
\item whether larger $k$ actually gives a dramatic slowdown in query speeds,
  which \S~\ref{sec:slowquery-theory} predicted was 
possible (\S~\ref{sec:queries-exper});
\item whether word length has a significant effect on the performance of
 EWAH (\S~\ref{sec:wordlen-exper});
\item whether 64-bit indexes are faster than 32-bit index when  aggregating many bitmaps (\S~\ref{sec:rangequeries-exper}).
\end{itemize}

In all of our experiment involving
32-bit words (our usual case), we choose to implement
EWAH with 16-bit counters to compress clean words.
When there are runs with  many more than $2^{16}$~clean words,
32-bit EWAH might be inefficient. However, on
our data sets, no more than 14\% of all counters
had the maximal value on sorted indexes, and no more than 3\% on 
unsorted indexes(see Table~\ref{tab:percentageOf32Overruns}).
Hence, EWAH is less efficient
 than WAH by a factor of no more than 14\% at storing the clean words.
 However, EWAH is more efficient than WAH by a constant factor of 3\% at
  storing the dirty words.
The last column in Table~\ref{tab:percentageOf32Overruns} shows
 runs of clean words make up only about half the storage;
 the rest is made of dirty words. 
 For 64-bit indexes, we have not seen any overrun.
 
\begin{table}
%

     \caption{Percentage of overruns in clean word compression
     using 32-bit EWAH with unary bitmaps and lexicographically sorted tables }\label{tab:percentageOf32Overruns}
    \centering
    \subtable[lexicographically sorted]{
    \begin{tabular}{l|rr|}\cline{2-3}
       & overruns & {\scriptsize$\frac{\mathrm{clean\ runs}}{\mathrm{total\ size}}$} \\ \hline
         \multicolumn{1}{|l|}{\textbf{Census-Income} (4-d)}      &  0\% & 60\% \\
         \multicolumn{1}{|l|}{\textbf{DBGEN} (4-d)}      &  13\% & 44\%  \\
         \multicolumn{1}{|l|}{\textbf{Netflix}}      &  14\%  & 49\%   \\
         \multicolumn{1}{|l|}{\textbf{KJV-4grams}}       &   4.3\%  & 43\%   \\
\hline
    \end{tabular}
    }
    \subtable[unsorted]{
    \begin{tabular}{l|rr|}\cline{2-3}
       & overruns & {\scriptsize$\frac{\mathrm{clean\ runs}}{\mathrm{total\ size}}$} \\ \hline
         \multicolumn{1}{|l|}{\textbf{Census-Income} (4-d)}      &  0\% & 52\% \\
         \multicolumn{1}{|l|}{\textbf{DBGEN} (4-d)}      &  0.2\% & 45\%  \\
         \multicolumn{1}{|l|}{\textbf{Netflix}}      &  2.4\%  & 49\%   \\
         \multicolumn{1}{|l|}{\textbf{KJV-4grams}}       &   0.1\%  & 47\%   \\
\hline
    \end{tabular}
    }

\end{table}






\subsection{Sorting disjoint blocks}
\label{subsection:crazyblocks}
Instead of sorting the entire table, we may 
partition the table horizontally into 
disjoint blocks. Each block can then be sorted
lexicographically and the table reconstructed.
Given $B$~blocks, the sorting complexity goes
from $O(n\log n)$ to $O(n \log n/B)$.
Furthermore, if blocks are small enough, we can sort in main memory.
Unfortunately, the indexing time and bitmap sizes both substantially increase,
even with only 5~blocks. (See Table~\ref{tab:SortBlockTime}.)
Altogether, sorting by blocks does not seem useful.

Hence, competitive row-reordering
alternatives 
should be scalable to
a large number of rows. For example, any 
heuristic in $\Omega(n^2)$ is probably irrelevant.

\begin{table}
\centering
\caption{Time required to sort and index, and sum of the compressed sizes of the bitmaps, for 
 $k=1$ (time in seconds and size in MB). Only three columns of each data sets are used with cardinalities of 7, 11, 400\,000  for DBGEN and of 5, 2\,182 and 17\,770 for Netflix. 
}

\begin{tabular}{|c|ccccc|ccccc|} \cline{2-6}
 \multicolumn{1}{c}{} &  \multicolumn{5}{|c|}{\textbf{DBGEN (3d)}} \\ \hline 
\# of blocks	& sort & fusion  & indexing  &total  & size  \\ \hline  
1 (complete sort) & 31 & - &  65 & 96 & 39 \\
5 & 28 & 2 & 68 & 98  & 51 \\ 
10 & 24 & 3 & 70 & 99  & 58  \\ 
500 & 17 & 3 & 87 & 107 & 116 \\ 
no sorting   & -  & -  & 100 & 100 & 119  \\\hline
 \multicolumn{1}{c}{} & \multicolumn{5}{|c|}{\textbf{Netflix (3d)}} \\ \hline 
1 (complete sort) & 487 & - & 558 & 1\,045  & 129\\
5   &	360	& 85 & 572 & 1\,017  & 264  \\ 
10 & 326 & 87 & 575 & 986   & 318 \\ 
500 &  230 & 86 & 601 & 917  & 806 \\ 
no sorting   & - & - & 689 & 689  & 1\,552 \\\hline
\end{tabular}
\label{tab:SortBlockTime}
\end{table}

\subsection{Index construction time}
\label{subsection:IndexConstructionTime}

Table~\ref{tab:SortBlockTime} shows that sorting may increase
the overall index-construction time (by 35\% for Netflix).
While Netflix and DBGEN nearly fit in the machine's main memory (2\,GiB), KJV-4grams is much larger (21.6\,GB).
Constructing a simple bitmap index (using Gray-Lex) over KJV-4grams took
approximately 14\,000~s or less than four hours.  
Nearly half (6\,000\,s) of the time 
was due to the \textbf{sort} utility, since the data set exceeds 
the machine's main memory (21.6\,GB vs. 2\,GiB).  Constructing an unsorted 
index is faster (approximately 10\,000\,s or 30\% less), but the index is about 9~times
larger (see Table~\ref{tab:index-sizes}).  

For DBGEN, Netflix and KJV-4grams,  
the construction of the bitmap index itself over the sorted table
is faster by at least 20\%. This effect is so significant over
DBGEN that it is faster to first sort prior to indexing.

\subsection{Sorting}
\label{sec:exper-sorting}
On some synthetic Zipfian data sets, we found
a small improvement 
(less than 4\% for 2~dimensions) by using Gray-Lex 
 in preference
to Binary-Lex. 
\begin{figure*} 
\centering
 \subfigure[Binary-lex versus shuffle]{\includegraphics[height=0.48\textwidth, angle=270]{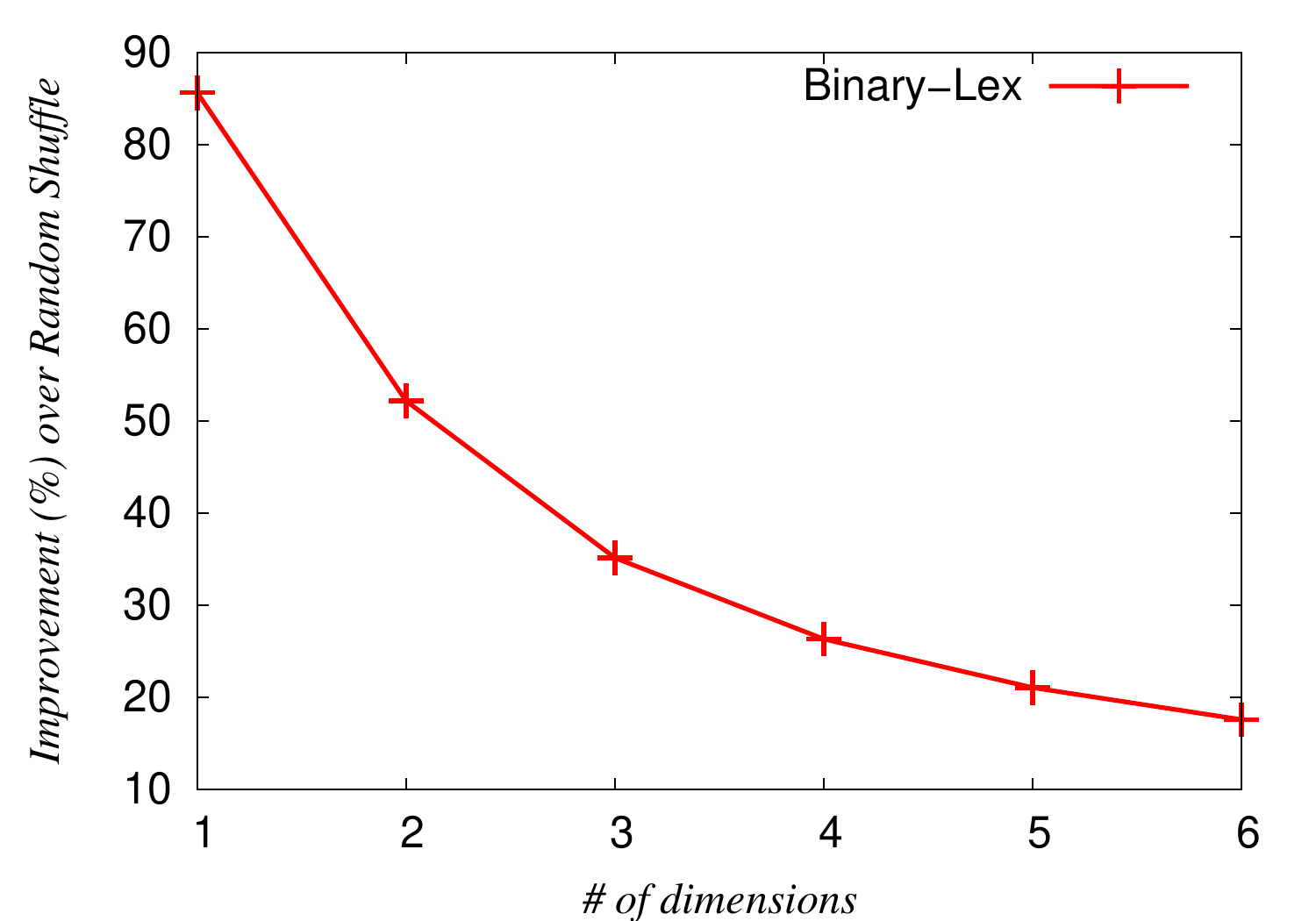}\label{fig:zipf-ktwothree-rands}}
 \subfigure[Gray-lex versus binary-lex 
]{\includegraphics[height=0.48\textwidth, angle=270]{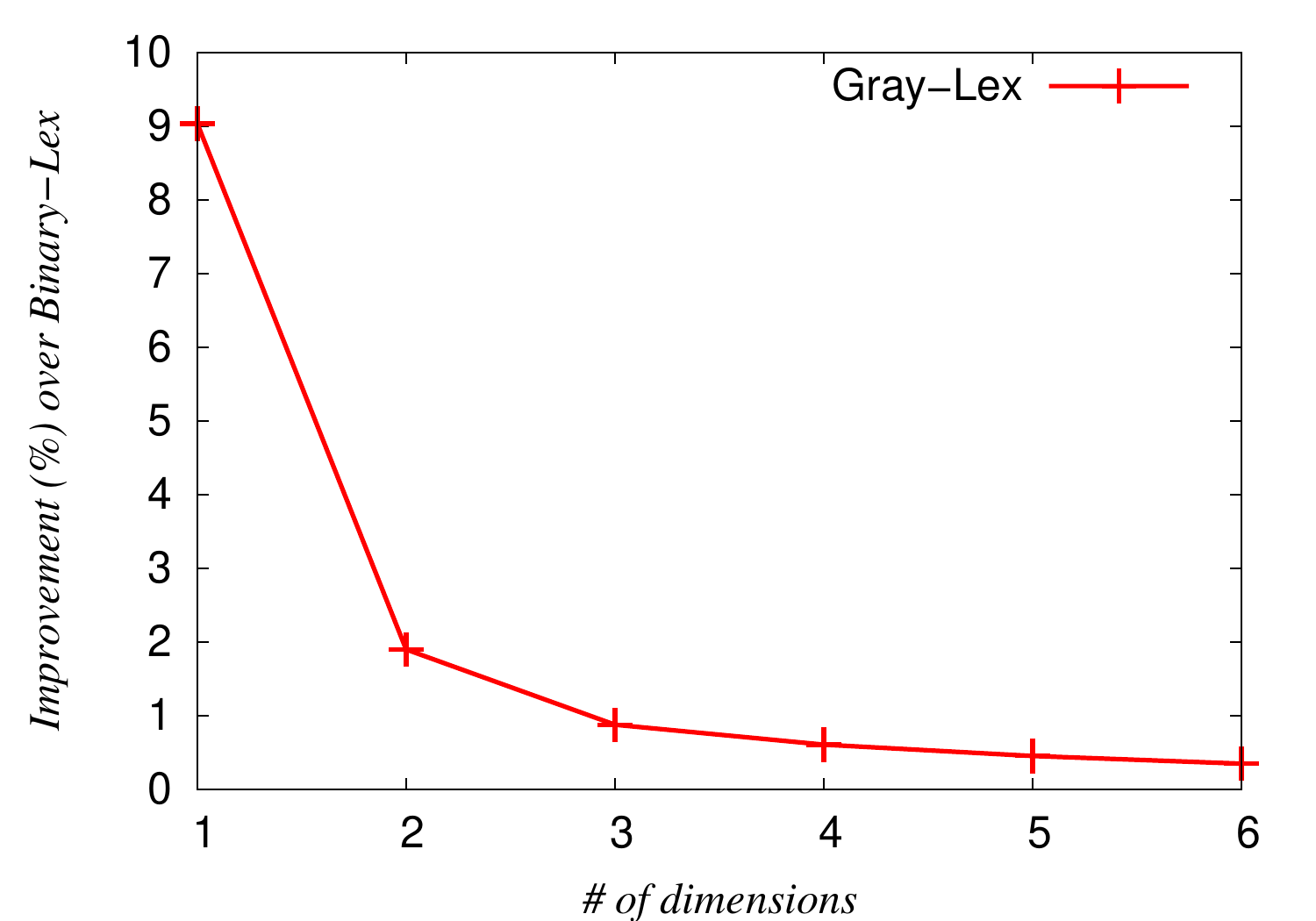}\label{fig:zipf-kthree}}
\caption{\label{fig:relative-perf-zipf}
  Relative performance, as a function of the number of
dimensions, on a Zipfian data set. 
}
\end{figure*}
Our data sets have 100~attribute values per dimension, and the frequency
of the attribute values is Zipfian (proportional to $1/r$, where $r$ is
the rank of an item). Dimensions were independent of one another.
See Fig.~\ref{fig:relative-perf-zipf}, where we compare Binary-Lex
to an unsorted table, and then Gray-Lex to Binary-Lex. For the latter,
the advantage drops quickly with the number of dimensions.
For one dimension, the performance improvement is 9\% for $k=2$, but for
more than 2 dimensions, it is less than 2\%.
On other data sets, Gray-Lex
either had no effect or a small positive effect.

\begin{table}
\caption{ \label{tab:index-sizes} Total sizes (words) of 32-bit EWAH bitmaps for various sorting methods. 
}
\centering\small
\begin{tabular}{|cr|rrrrr|}\cline{2-7} 
\multicolumn{1}{c|}{} & 
\multicolumn{1}{c|}{k}& Unsorted   & Rand-Lex                             & Binary-Lex                             & Gray-Lex                               & Gray-Freq.                       \killforgood{& Freq-Component} \\ \hline
Census-       & 1   & $8.49\times10^5$ &\killforgood{  487319}$4.87\times10^5$  &\killforgood{  487319}$4.87\times10^5$ &\killforgood{  487319  }$4.87\times10^5$& \killforgood{487464   }$4.87\times10^5$\killforgood{& 617381}    \\
Income  &      2    & $9.12\times10^5$   &\killforgood{  652689}$6.53\times10^5$  &\killforgood{  452635}$4.53\times10^5$ &\killforgood{  451927  }$4.52\times10^5$& \killforgood{435595   }$4.36\times10^5$\killforgood{& 426312}    \\
(4d)    &      3    & $6.90\times10^5$   &\killforgood{  485368}$4.85\times10^5$  &\killforgood{  376812}$3.77\times10^5$ &\killforgood{  372728  }$3.73\times10^5$& \killforgood{327939   }$3.28\times10^5$\killforgood{& 377165}    \\
      &      4    & $4.58\times10^5$   &\killforgood{  274200}$2.74\times10^5$  &\killforgood{  222670}$2.23\times10^5$ &\killforgood{  217397  }$2.17\times10^5$& \killforgood{197695   }$1.98\times10^5$\killforgood{& 329624}    \\ \hline
DBGEN &      1    &  $5.48\times10^7$  &\killforgood{33847417}$3.38\times10^7$  &\killforgood{33847417}$3.38\times10^7$ &\killforgood{  33847445}$3.38\times10^7$& \killforgood{33847116 }$3.38\times10^7$\killforgood{&  33851895}     \\
(4d)  &      2    &  $7.13\times10^7$  &\killforgood{28972175}$2.90\times10^7$  &\killforgood{27647998}$2.76\times10^7$ &\killforgood{27639093  }$2.76\times10^7$& \killforgood{27441584 }$2.74\times10^7$\killforgood{&  27816958}     \\
      &      3    &  $5.25\times10^7$  &\killforgood{17271023}$1.73\times10^7$  &\killforgood{15064193}$1.51\times10^7$ &\killforgood{15038604  }$1.50\times10^7$& \killforgood{14977582 }$1.50\times10^7$\killforgood{&  15542383}     \\
      &      4    &   $3.24\times10^7$ &\killforgood{15171557}$1.52\times10^7$  &\killforgood{12134337}$1.21\times10^7$ &\killforgood{ 12060794 }$1.21\times10^7$& \killforgood{11906432 }$1.19\times10^7$\killforgood{&  12877534}     \\ \hline
Netflix &    1    &   $6.20\times10^8$ &\killforgood{321829546}$3.22\times10^8$ &\killforgood{321829546}$3.22\times10^8$&\killforgood{321829546 }$3.22\times10^8$& \killforgood{318959993}$3.19\times10^8$\killforgood{& 313900325 }    \\
      &      2    &   $8.27\times10^8$ &\killforgood{417747399}$4.18\times10^8$ &\killforgood{316661579}$3.17\times10^8$&\killforgood{316826482 }$3.17\times10^8$& \killforgood{242676792}$2.43\times10^8$\killforgood{& 388117623 }    \\
      &      3    &   $5.73\times10^8$ &\killforgood{240247295}$2.40\times10^8$ &\killforgood{198127358}$1.98\times10^8$&\killforgood{196575343 }$1.97\times10^8$& \killforgood{149062457}$1.49\times10^8$\killforgood{& 226287832 }    \\
      &      4    &   $3.42\times10^8$ &\killforgood{160288113}$1.60\times10^8$ &\killforgood{138656126}$1.39\times10^8$&\killforgood{137067789 }$1.37\times10^8$& \killforgood{113820043}$1.14\times10^8$\killforgood{& 152122246 }    \\ \hline
KJV-  &      1    & $6.08\times10^9$ &\killforgood{668244925}$6.68\times10^8$ &\killforgood{668244925}$6.68\times10^8$&\killforgood{ 668244925}$6.68\times10^8$& \killforgood{667570961}$6.68\times10^8$\killforgood{& n/a       }    \\
4grams&      2    &  $8.02\times10^9$  &\killforgood{1088137352}$1.09\times10^9$&\killforgood{1005828896}$1.01\times10^9$&\killforgood{992824611}$9.93\times10^8$& \killforgood{729015073}$7.29\times10^8$\killforgood{& n/a       }    \\
      &      3    & $4.13\times10^9$   &\killforgood{920400236}$9.20\times10^8$ &\killforgood{833869069}$8.34\times10^8$&\killforgood{ 830621456}$8.31\times10^8$& \killforgood{577319979}$5.77\times10^8$\killforgood{& n/a       }    \\
      &      4    & $2.52\times10^9$   &\killforgood{723455063}$7.23\times10^8$ &\killforgood{648552452}$6.49\times10^8$&\killforgood{ 638934508}$6.39\times10^8$& \killforgood{500834200}$5.01\times10^8$\killforgood{& n/a       }    \\ \hline
\end{tabular}
\end{table}

Table~\ref{tab:index-sizes} shows the sum of bitmap sizes 
using Gray-Lex orderings
and Gray-Frequency.
For comparison, we also used an unsorted table (the code allocation should
not matter; we used the same code allocation as Binary-Lex), and we
used a random code assignment with a lexicographically sorted table (\emph{Rand-Lex}).
Dimensions were ordered from the largest to the smallest (``4321'')
except for Census-Income where we used the ordering ``3214''.

KJV-4grams had a larger index for  $k=2$ than $k=1$.
This data set has many very long runs of identical attribute values
in the first two dimensions, and the number of
attribute values is modest, compared with the number of rows.
This is ideal for 1-of-$N$.


For $k=1$, as expected,  encoding is irrelevant: Rand-Lex, Binary-Lex, Gray-Lex, and Gray-Freq have identical results. However, sorting the table lexicographically is important: the reduction in size of the bitmaps is about 40\%
for 3~data sets (Census-Income, DBGEN, Netflix),  and goes up to  90\% 
for KJV-4grams.

For $k>1$,
Gray-Frequency yields the  smallest indexes in Table~\ref{tab:index-sizes}. The difference with the second-best, Gray-Lex, can be substantial (25\%) but is typically small. However, Gray-Frequency is histogram-aware and thus, more complex to implement. The difference between Gray-Lex and Binary-Lex is small even though Gray-Lex is sometimes slightly better ($\approx$2\%) especially for denser indexes ($k=4$). However, Rand-Lex is noticeably worse (up to $\approx$25\%) than both of them: this means that encoding is a significant issue. All three schemes (Binary-Lex, Gray-Lex, Rand-Lex) have about the same complexity---all three are histogram-oblivious---and therefore Gray-Lex is recommended.

We omit Frequent-Component from the table.  On Netflix, for $k=1$, it
outperformed the other approaches by 1\%, and for DBGEN it was only slightly
worse than the others. But in all other case on DBGEN, Census-Income and Netflix,
it lead to indexes 5--50\% larger.
(For instance, on Netflix ($k=4$) the index size was $1.52 \times 10^8$ words,
barely better than Rand-Lex and substantially worse than Gray-Frequency.)
Because it interleaves attribute values and it is histogram-aware, it may be the most difficult scheme to implement efficiently among our candidates. Hence, we recommend against Frequent-Component.

\subsection{Column effects}
\label{sec:column-exper}
We experimentally evaluated how lexicographic sorting affects the EWAH
compression of individual columns.  Whereas
sorting tends to create runs of identical values in the first
columns,  the benefits of sorting are far less
apparent in later columns, except those strongly
correlated with the first few columns.  
\killforgood{(As a thought experiment,
consider a table with 100 columns, 99 of which are identical.
An ordering which is moderately helpful to any one of the 99 columns
would probably be preferable to an ordering that is substantially
helpful to the remaining column.)}
For Table~\ref{tab:sizesortcolumnorderd10},
 we have sorted
projections of Census-Income and DBGEN onto 10~dimensions
$d_1\ldots{}d_{10}$ with $n_1<\ldots<n_{10}$.  
(The
dimensions $d_1 \ldots d_4$ in this group are different from
the dimensions $d_1 \ldots d_4$ discussed earlier.)
We see that if we sort from the largest column ($d_{10}\ldots{}d_1$),
at most 3 columns benefit from the sort, whereas 5 or more columns benefit
when sorting from the smallest column ($d_1\ldots{}d_{10}$).
{%
\begin{table}
\caption{Number of 32-bit words used for different unary indexes when the
table was sorted lexicographically (dimensions ordered
by descending cardinality, $d_{10}\ldots d_1$, or by ascending
cardinality, $d_1\ldots d_{10}$).
}\label{tab:sizesortcolumnorderd10}
\centering
\subtable[Census-Income]{
\begin{tabular}{|c|rrrr|}\cline{2-5}
	\multicolumn{1}{c|}{} & cardinality & unsorted & \multicolumn{1}{c}{$d_1 \ldots d_{10}$} & \multicolumn{1}{c|}{$d_{10} \ldots d_1$} \\ \hline
$d_1$    & 7   & 42\,427 & 32 & 42\,309 \\
$d_2$    & 8   & 36\,980 & 200 & 36\,521  \\
$d_3$    & 10  & 34\,257 & 1\,215 & 28\,975  \\
$d_4$    & 47  & 0.13$\times 10^6$ & 12\,118 & 0.13$\times 10^6$  \\
$d_5$    & 51  & 35\,203 & 17\,789 & 28\,803 \\
$d_6$    & 91  & 0.27$\times 10^6$ & 75\,065 & 0.25$\times 10^6$  \\
$d_7$    & 113 & 12\,199 &  9\,217 & 12\,178  \\
$d_8$    & 132 & 20\,028 & 14\,062 & 19\,917  \\
$d_9$    & 1\,240 & 29\,223 & 24\,313 & 28\,673  \\
$d_{10}$ & 99\,800 & 0.50$\times 10^6$ & 0.48$\times 10^6$ & 0.30$\times 10^6$  \\ \hline
total    & - & 1.11$\times 10^6$ & 0.64$\times 10^6$ & 0.87$\times 10^6$   \\ \hline
\end{tabular}
}
\subtable[DBGEN]{
\begin{tabular}{|c|rrrr|}\cline{2-5}
	\multicolumn{1}{c|}{} & cardinality & unsorted & \multicolumn{1}{c}{$d_1 \ldots d_{10}$} & \multicolumn{1}{c|}{$d_{10} \ldots d_1$} \\ \hline
$d_1$ & 2 & 0.75$\times 10^6$ & 24 & 0.75$\times 10^6$ \\
$d_2$ &  3 &  1.11$\times 10^6$ & 38 & 1.11$\times 10^6$ \\
$d_3$ & 7 & 2.58$\times 10^6$ & 150 & 2.78$\times 10^6$ \\
$d_4$ &  9 & 0.37$\times 10^6$ & 1\,006 & 3.37$\times 10^6$ \\
$d_5$ & 11 & 4.11$\times 10^6$ & 10\,824 & 4.11$\times 10^6$ \\
$d_6$ & 50 & 13.60$\times 10^6$ & 0.44$\times 10^6$ & 1.42$\times 10^6$ \\
$d_7$ & 2\,526 & 23.69$\times 10^6$ & 22.41$\times 10^6$ & 23.69$\times 10^6$ \\
$d_8$ & 20\'000 & 24.00$\times 10^6$ & 24.00$\times 10^6$ & 22.12$\times 10^6$ \\
$d_9$ & 400\,000 & 24.84$\times 10^6$ & 24.84$\times 10^6$ & 19.14$\times 10^6$ \\
$d_{10}$ & 984\,297 & 27.36$\times 10^6$ & 27.31$\times 10^6$ & 0.88$\times 10^6$ \\ \hline
total  & - & 0.122$\times 10^9$ & 0.099$\times 10^9$ & $0.079\times 10^9$ \\ \hline
\end{tabular}
}
\end{table}
}

\begin{figure*}
\centering
  \subfigure[Census-Income]{\includegraphics[width=0.48\textwidth]{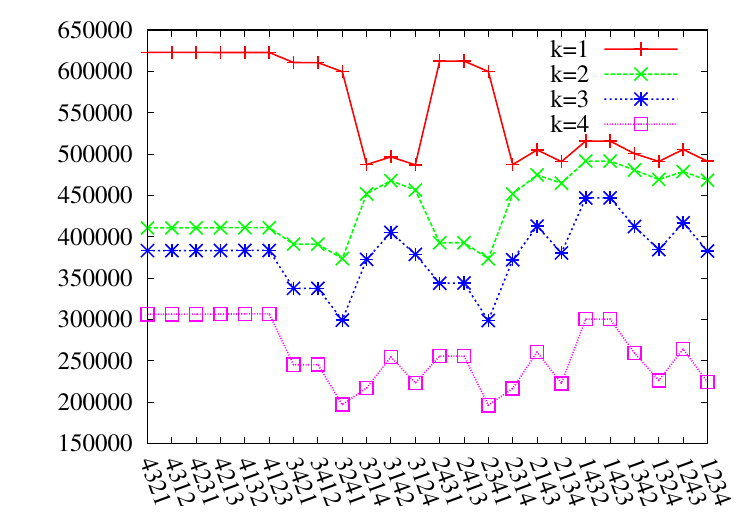}\label{fig:census-order}}
 \subfigure[DBGEN]{\includegraphics[width=0.48\textwidth]{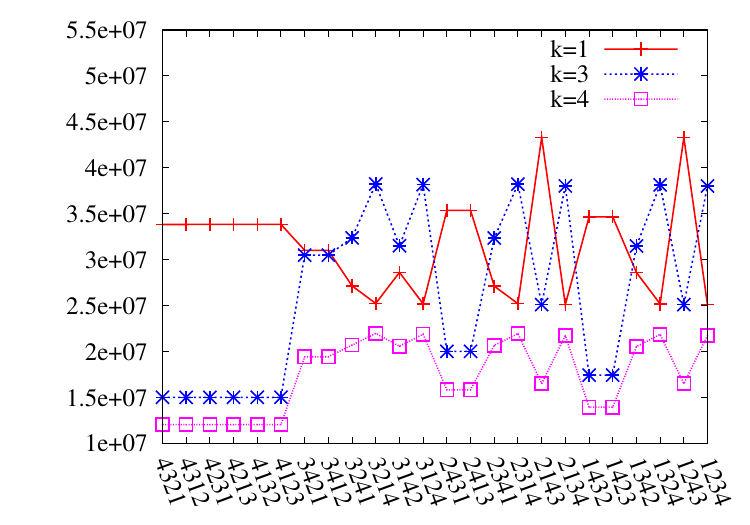}\label{fig:dbgen-orders}}
  \subfigure[Netflix]{\includegraphics[width=0.48\textwidth]{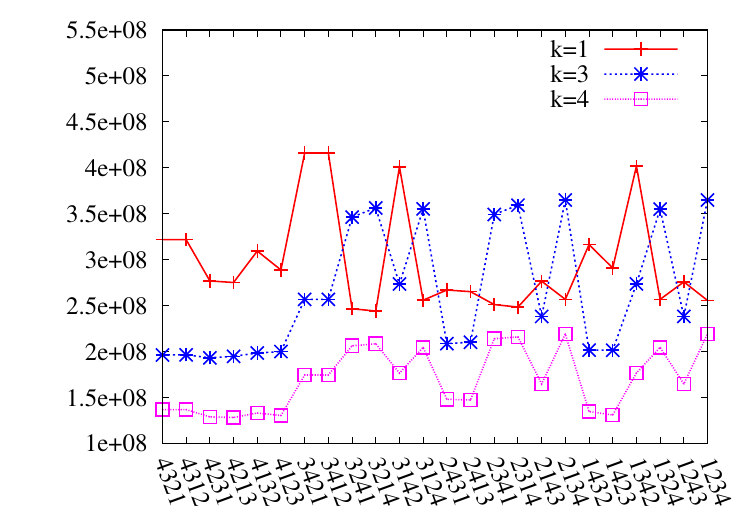}\label{fig:nf-order}}
\caption{Sum of EWAH bitmap  sizes (words, $y$ axis) on 4-d data sets 
for all dimension orderings ($x$ axis).
\killforgood{now explanation has to come earlier
Ordering $abcd$ indicates the $a^{\mathrm{th}}$ smallest dimension's
being the primary sort key, the $b^{\mathrm{th}}$ smallest dimension's
being the secondary sort key, and so forth.}} \label{fig:orders}
\end{figure*}

We also assessed how the total size of the index was
affected by various column orderings;
we show  
 the Gray-Lex index
sizes for each column ordering in Fig.~\ref{fig:orders}. 
The dimensions
of KJV-4grams are too similar for ordering to be interesting,
\altnotessential{and we have  omitted them. }%
\killforgood{and thus we 
used TWEED instead in our tests.  
\daniel{See above remark regarding why TWEED might be evil.}
\owen{If we can explain this, it is a good contribution.
It shows why toy data sets are misleading.  Plus, some people will
want to build bitmap indexes on small relations and we would not
want to mislead them.}
\owen{Another reason for wanting TWEED is that we cannot nicely
show 3 data sets.  For 2, we need an excuse why we don't show the
3rd (what are we hiding?)  We could revive DBLP, though}}%
 For small dimensions, the value of $k$ was lowered using
the heuristic presented in \S~\ref{sec:bitmapIndexes}.
Our results suggest that table-column reordering has a 
significant effect (40\%). 
\killforgood{This does not directly 
contradict the observation
by Canahuate et.\ al~\cite{pinarunpublished} that
bitmap
 reordering 
 does not change the 
 size much. }\killforgood{did not seem to be important, because they were
reordering bitmaps after encoding, rather than attributes that are
then encoded into an index.}%

The value of $k$ affects which ordering leads to
the smallest index: 
good orderings for  $k=1$ are frequently bad orderings for
$k>1$, and vice versa.
This is consistent with our earlier analysis (see Figs.~\ref{fig:theorygain} and~\ref{fig:toyorderings}).
\killforgood{For TWEED, all values
of $k$ suggested similar orderings (for instance, the second
largest dimension should not go first), although they differed
somewhat about which orders were best, if the third-largest
dimension goes first. 
\daniel{This behavior of TWEED is consistent with what my theoretical
investigations reveal. I say we should drop TWEED and explain that
we are interested in large tables only.}
However, for the other data sets, 
}%
For Netflix and DBGEN, we have omitted $k=2$ for legibility.


 Census-Income's largest dimension is very large ($n_4 \approx n/2$);
 DBGEN also has a large dimension ($n_4 \approx n/35$). Sorting columns
 in decreasing order with respect to  $\min(n_i^{-1/k}, (1-n_i^{-1/k})/(4w-1))$ for $k=1$,
 we have that only for DBGEN   the ordering
 ``2134'' is suggested, otherwise,  ``1234'' (from smallest to largest) is recommended. Thus the heuristic
 provides nearly optimal recommendations. For $k=3$ and $k=4$, the ordering ``1234'' is recommended for
 all data sets: for $k=4$ and Census-Income, this recommendation is wrong.
 For $k=2$ and Census-Income, the ordering ``3214'' is recommended, another wrong 
 recommendation for this data set. Hence, a better column reordering heuristic is needed for $k>1$.
 Our greedy approach may be too simple, and
it may be necessary to know the histogram skews.

\killforgood{%
\owen{This discussion can be replaced by the table?}
For TWEED, this is
not the best ordering for any value of $k$, but it is never much
worse than the optimal ordering.   In TWEED, it is always a bad
idea for $d_3$ (Year) to go first.  This dimension's histogram 
\cut{is
found in Fig.~\ref{fig:tweed-year-histo}; we see that there
are}
has
a moderately large number of values, none of which are overwhelmingly
common.  In contrast, the numerically larger dimension $d_4$ has a 
much more skewed histogram and the top 4 values account for more
than half the records.
\cut{histogram shown in Fig.~\ref{fig:tweed-agent-histo}.}  
For Census-Income, the suggested
order is good for $k=1$, but it is a poor choice for $k>1$, where
either ordering 3241 or 2341 would be better.  The overall idea to
avoid orderings with the excessively large dimension ($d_4$) first is good,
however: the first 6 orderings for Census-Income are poor choices for any $k$.
For DBGEN, we hide $k=2$, because for most orderings it consumed
more space than $k=1$.  
For Netflix, the suggested ordering is good (but not optimum) for
$k>1$.  However, it is a poor choice for $k=1$, when the opposite ordering
``1234'' would be a much better choice.


When we counted an attribute value only if it occurs more than 32 times,
the suggested ordering for Netflix and DBGEN is (still) ``4321,''
\notessential{for TWEED it is 3421 (although 4321 is a near second), }
for Census-Income is  1324.
This is a poor choice for \notessential{TWEED and} Netflix, a good choice for $k=1$ or
$k=4$ on Census-Index, but a poor choice for $k=2$ or $k=3$. 
For DBGEN, it is a good choice for $k>1$.

For $k=1$ on Netflix, we see that some orderings beginning with $d_3$
(specifically, 3214 and 3241)
are among the best, whereas  others (3412 and 3421) are especially bad.
Since dimension 1 (rating) is tiny, functionally dependent on
the  other dimension,  and there are many occurrences of each value,
it may be sensible to ignore this value.    Then, we see that ordering
324 is good, but 342 is bad.
\owen{badly needs an explanation.  Maybe, tuples of (3,2) are nicely 
distributed?}
DBGEN, a synthetic data set, performed well for $k>1$ as long as the
largest dimension was first: the ordering of the other dimensions did
not matter.  However, for $k=1$, the reversed ordering, ``1234,'' was
among the best.

\owen{Netflix k=1.  34xx is bad but 32xx is good.  Similarly, DBGEN 
1234/2143 good but 1243/2143 bad. Weird?}
}

\killforgood{
\subsection{Histogram-Oblivious Interleaving}

When tested, the scheme for interleaving bitmaps (see the end of 
\S~\ref{sec:dim-order}) worked poorly.
To avoid issues with from differing
sized dimensions, we experimented with synthetic cubes with
91 attributes per dimension distributed with Zipfian ($s=1.0$)
and 5000 rows (apprx 4000 distinct rows). 
A Gray-Lex bitmap index was generated, then bitmaps
were interleaved, and finally the result was GC sorted.

For $k=3$ and $d=3$, interleaving increased the size of
the bitmaps by an average of  5.4\% (taken over 1000 cubes). 
For $k=2$ and $d=3$, the average bitmap size increased
by 5.3\%.
\notessential{
On TWEED, we saw a 44\% size increase \owen{hmm which k}
from interleaving, compared with Gray-Lex with largest dimension
first.}
}
\killforgood{
The Frequent-Component approach seemed to work poorly 
(see Table~\ref{tab:index-sizes}).  Owen thinks he may
have run the tests incorrectly because he was not expecting this.
\daniel{I got the same type of results. Canahuate et al.~\cite{pinarunpublished}
reported that their interleaving was not very good at
increasing the compression, so we just confirm their results.
      the type of interleaving we do is akin to $k=1$ bitmap
reordering. There is no reason to believe it would work well
for $k>1$. So the fact that it does not make things too much
worse for $k=1$, but make things worse for $k>1$ should not
come as a surprise. }
}

\subsection{Index size growth}
\label{sec:exper-lexsorting}
\suggestwecut{%
Figs.~\ref{fig:kjv-size-vs-lines}~and~\ref{fig:kjv-time-vs-lines} 
show the sizes of the sum of unary bitmaps for various prefixes of KJV-4grams. 
It also shows the time to create these indexes.
\owen{It might be easier to grok as a log-log plot.  I ended up looking
at a nolog-nolog and log-log version while trying to understand it}
We see that the full index is nine times bigger if unsorted.  We also
see the sorted index appears to grow sublinearly, probably due to repeated
or almost repeated rows.  It should be faster to query a smaller index.
However, the time to sort and then create this smaller index is more than
the time to create the unsorted index.  The additional time is less than
the time to sort, and all times seem to scale linearly with the number
of rows. \daniel{We might stress here that this is external-memory sorting
since the machine has 2\,GiB of RAM\@ which is less than the size of the data set.}
}

To study scaling, we built indexes from prefixes
of the full KJV-4grams 
data set.  We found \killforgood{construction times} that the sum of the EWAH bitmap sizes (see Fig.~\ref{fig:kjv-size-vs-lines-gray}) increased linearly.
  Yet with sorting, the bitmap sizes increased sublinearly.
As new data arrives, it is increasingly likely to fit into existing
runs, once sorted.
Hence---everything else being equal---sorting becomes more beneficial as the data sets grow.

\suggestwecut{
	\begin{figure}
\centering
\includegraphics[width=.49\textwidth]{sizevsnboflines-alpha}
\caption{\label{fig:kjv-size-vs-lines} Sum of the EWAH bitmap sizes for various
prefixes of the KJV-4grams table, $k=1$, shuffled table
and lexicographic table order.}
\end{figure}
}

\suggestwecut{
\begin{figure}
\centering
\includegraphics[width=.55\textwidth]{timevsnboflines-alpha}
\caption{\label{fig:kjv-time-vs-lines} \owen{cpu or wall?}\daniel{Kamel used the "time" command for sort, but we don't have
the script to tell how he measured the indexing time} Times to create an index,
for various prefixes of the KJV-4grams table. $k=1$}
\end{figure}
}

\begin{figure}
\centering
\includegraphics[width=.49\textwidth]{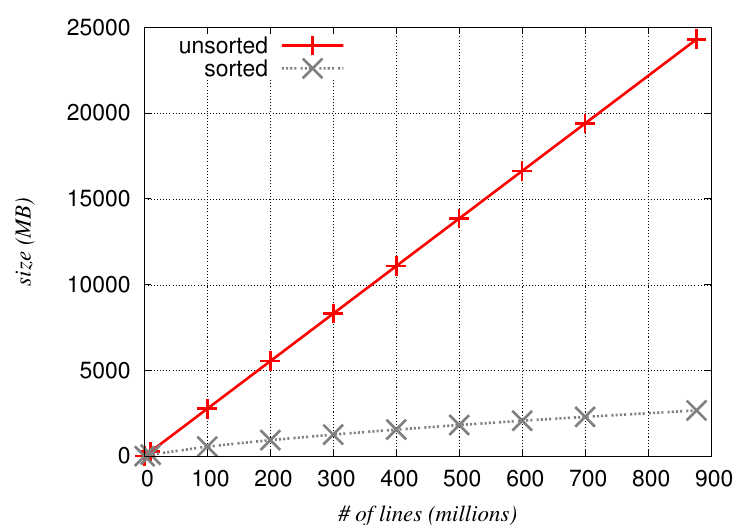}
\caption{\label{fig:kjv-size-vs-lines-gray} Sum of the EWAH bitmap sizes for various
prefixes of the KJV-4grams table ($k=1$)}
\end{figure}

\subsection{Bitmap reordering}
\label{sec:bitmap-reordering-exper} 
%

Sharma and Goyal~\cite{sharma2008emc} consider encoding a 
table into a bitmap index using a multi-component code (similar
to $k$-of-$N$), then GC sorting the rows of the index, and finally
applying WAH compression. 
Canahuate et al.~\cite{pinarunpublished} propose a similar
approach, with the additional step of permuting
the columns---meaning the individual bitmaps---in the index prior to GC sorting. 
For example, whereas the list of 2-of-4 codes in increasing GC
order is
 0011, 0110, 0101, 1100, 1010, 1001, by permuting the first and
 the last bit, we obtain the following  (non-standard) Gray code:
  1010, 0110, 1100, 0101, 0011, 1001.
{In effect, reordering bitmaps is equivalent to sorting the (unpermuted) index rows
according to a non-standard Gray code.} 
They chose to use bitmap
density to determine which index columns should come first, but
reported that the different orders had little effect on the
final index sizes. 
\killforgood{
\daniel{Here owen correctly reported
that Canahuate et al. say something about generating uniform
bitmap sizes. But that has little to do with the ongoing 
discussion, and it is a rather fine point.}
\owen{if we we get rejected, we can reconsider this one.}
\suggestwecut{ However, they did report making
the compressed bitmaps more uniform in size, thereby
reducing the potential for expensive queries.}
}

In contrast,
our approach has been to permute the columns of the table---not the individual bitmaps,
then sort the table lexicographically, and finally generate the compressed index. 
Permuting the attributes corresponds to permuting \emph{blocks} of bitmaps:
our bitmap permutations are a special case of Canahaute's. 
We do not know a sufficiently efficient method to sort our largest
data sets with arbitrary bitmap reordering.  
We cannot construct 
the uncompressed index: for KJV-4grams, we would require at least 3.7\,TB\@.   Instead, we used the compressed
B-tree approach mentioned in \S~\ref{sec:sorting-rows} and applied the
bitmap permutation to its keys.  This was about 100 
times slower than our normal Gray-Lex method, and implementation restrictions
prevented our processing the full Netflix or KJV-4grams data sets.
Hence, we took the first 20~million records from each of these two
data sets, forming Netflix20M and KJV20M.

Our experiments showed that little compression was lost 
by restricting ourselves to the special case of permuting table columns, rather than individual bitmaps.
While we indexed all the $4!=24$~tables generated by 
all column permutations in our 4-column data sets, 
it is infeasible to consider all bitmap
permutations. Even if there were only 100~bitmaps,
the number of permutations would be prohibitively large ($100!\approx 10^{158}$).
We considered three heuristics  based on bitmap density $\cal D$---the number of 1-bits over the total number of bits ($n$) 
: 
\begin{enumerate}
\item ``Incompressible first'' (IF), which orders bitmaps by
increasing $|\, {\cal D} - 0.5\,|$.  In other words, bitmaps with density
near 0.5 are first~\cite{pinarunpublished}. 
\item ``Moderately sparse first'' (MSF), ordering
by the value $\min({\cal D},\frac{1-{\cal D}}{4\times 32 - 1})$
as discussed at the end of \S~\ref{sec:dimOrder}. This is a per-bitmap variant of the column-reordering heuristic we evaluate experimentally in \S~\ref{sec:column-exper}.
\item ``Sparse first'' (SF): order by increasing ${\cal D}$.
\end{enumerate}
%
Results are shown in Table~\ref{tab:per-bitmap}.  In only one case
(KJV20M, $k=1$), was a
per-bitmap result significantly  better (by 5\%) than our default method of
rearranging table columns instead of individual bitmaps. 
In most other cases, all per-bitmap reorderings were worse,
 sometimes by large factors (30\%). 

IF
ordering performs poorly when there are some dense bitmaps (i.e., when
$k>1$.) Likewise, SF performs poorly for sparse bitmaps ($k=1$).
 We do not confirm prior reports~\cite{pinarunpublished} that
index column order has relatively little effect on the index size: on
our data, it makes a substantial difference.
Perhaps the characteristics of their scientific data sets
account for this difference.

\begin{table}
\caption{ \label{tab:per-bitmap} Sum of the EWAH bitmap sizes (in words), GC sorting
and various bitmap orders}
\centering\small
\begin{tabular}{|cr|c|ccc|}\cline{3-6}
\multicolumn{2}{c|}{}     & Best column        &  \multicolumn{3}{c|}{Per-bitmap reordering} \\
\multicolumn{2}{c|}{}     &             order         & IF               &  MSF             & SF \\ \hline
Census-Income &  $k=1$    & $4.87\times 10^{5}$  & $4.91\times 10^5$ & $4.91\times10^5$ & $6.18\times10^5$\\
(4d)          &      2    & $3.74\times 10^{5}$  & $4.69\times 10^5$ & $4.10\times10^5$ & $3.97\times10^5$\\
              &      3    & $2.99\times 10^{5}$  & $3.83\times 10^5$ & $3.00\times10^5$ & $3.77\times10^5$\\
              &      4    & $1.96\times 10^{5}$  & $3.02\times 10^5$ & $\mathbf{1.91\times10^5}$ & $\mathbf{1.91\times10^5}$\\\hline
\rule{0mm}{1em}DBGEN         
              &      1    & $2.51\times 10^{7}$  & $2.51\times 10^7$ & $2.51\times10^7$ & $3.39\times10^7$\\
(4d)          &      2    & $2.76\times 10^{7}$  & $4.50\times 10^7$ & $4.35\times10^7$ & $2.76\times10^7$\\
              &      3    & $1.50\times 10^{7}$  & $3.80\times 10^7$ & $1.50\times10^7$ & $1.50\times10^7$\\
              &      4    & $1.21\times 10^{7}$  & $2.18\times 10^7$ & $1.21\times10^7$ & $1.21\times10^7$\\\hline
\rule{0mm}{1em}Netflix20M    
              &      1    & $5.48\times 10^{7}$  & $5.87\times 10^7$ & $5.87\times10^7$ & $6.63\times10^7$\\
              &      2    & $7.62\times 10^{7}$  & $9.05\times 10^7$ & $8.61\times10^7$ & $7.64\times10^7$\\
              &      3    & $4.43\times 10^{7}$  & $7.99\times 10^7$ & $\mathbf{4.39\times10^7}$ & $\mathbf{4.39\times10^7}$\\
              &      4    & $2.99\times 10^{7}$  & $4.82\times 10^7$ & $3.00\times10^7$ & $3.00\times10^7$\\\hline
\rule{0mm}{1em}KJV20M        
              &      1    & $4.06\times 10^{7}$  & $4.85\times 10^7$ & $4.83\times10^7$ & $\mathbf{3.85\times10^7}$\\
              &      2    & $5.77\times 10^{7}$  & $6.46\times 10^7$ & $\mathbf{5.73\times10^7}$ & $\mathbf{5.73\times10^7}$\\
              &      3    & $3.95\times 10^{7}$  & $4.47\times 10^7$ & $4.24\times10^7$ & $4.24\times10^7$\\
              &      4    & $2.72\times 10^{7}$  & $3.42\times 10^7$ & $3.38\times10^7$ & $3.38\times10^7$\\\hline
\end{tabular}
\end{table}

\subsection{Queries}
\label{sec:queries-exper}

%
We implemented queries over the bitmap indexes by processing the logical operations two bitmaps at a time: we did not use
Algorithm~\ref{algo:genrunlengthmultiplefaster}.
Bitmaps are processed  in sequential order, without sorting by
size, for example. The query processing costs includes the extraction of the row IDs---the location of the 1-bits---from the bitmap form of the result.

\begin{figure*}
\centering
%
 \subfigure[Unsorted table]{\includegraphics[width=0.49\columnwidth]{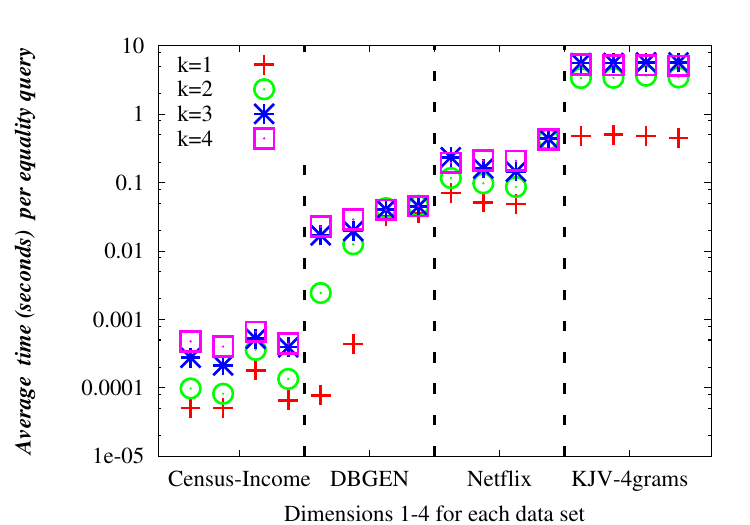}\label{fig:randqueries-times-unsorted}}
 \subfigure[Sorted table and Gray-Lex encoding]{\includegraphics[width=0.49\columnwidth]{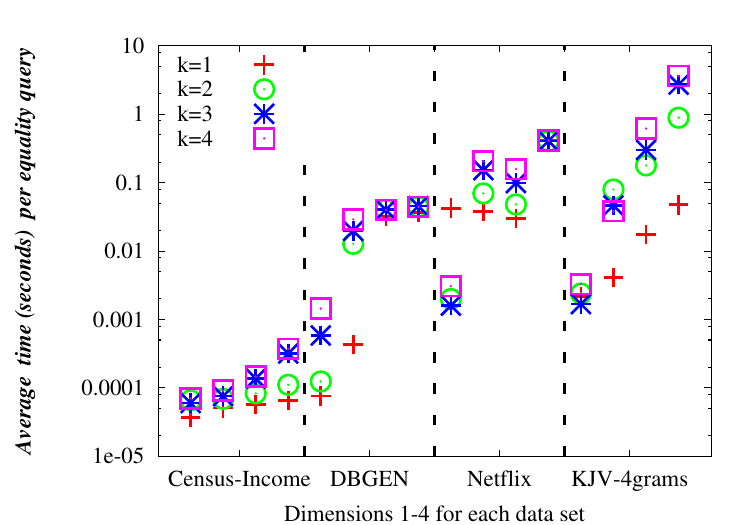}\label{fig:randqueries-times}}
\caption{Query times are affected by dimension, table sorting and $k$.
}\label{fig:queries}
\end{figure*}

We timed equality queries against our 4-d bitmap indexes.
Recall that dimensions were ordered from the largest to the smallest (4321) 
 except for Census-Income where 
 we used the ordering ``3214.''
Gray-Lex encoding is used for $k>1$.
 Queries were generated by choosing attribute values uniformly
at random and the figures report average wall-clock times for such queries.
We made 100 random choices per column
for KJV-4grams when $k>1$.
For DBGEN and Netflix, we had 1\,000 random choices per column
and 10\,000 random choices were used for Census-Income
and KJV-4grams ($k=1$). For each  data set, 
we give the results per column (leftmost tick is the 
column used as the primary sort key, next tick is for the
secondary sort key, etc.).
The
results are shown in Fig.~\ref{fig:queries}.

From Fig.~\ref{fig:randqueries-times}, we see that simple bitmap
indexes almost always yield the fastest queries.  The difference caused
by $k$ is highly dependent upon the data set and the particular column
in the data set. However, for a given data set and column, with only
a few small exceptions, query times increase with $k$, especially from
$k=1$ to $k=2$.
For DBGEN, the last two dimensions have size 7 and 11, whereas
for Netflix, the last dimension has size 5, and therefore,
they will never use a $k$-value larger than 2: their speed is mostly
oblivious to  $k$.

\begin{figure} 
\centering
{\includegraphics[width=0.49\columnwidth]{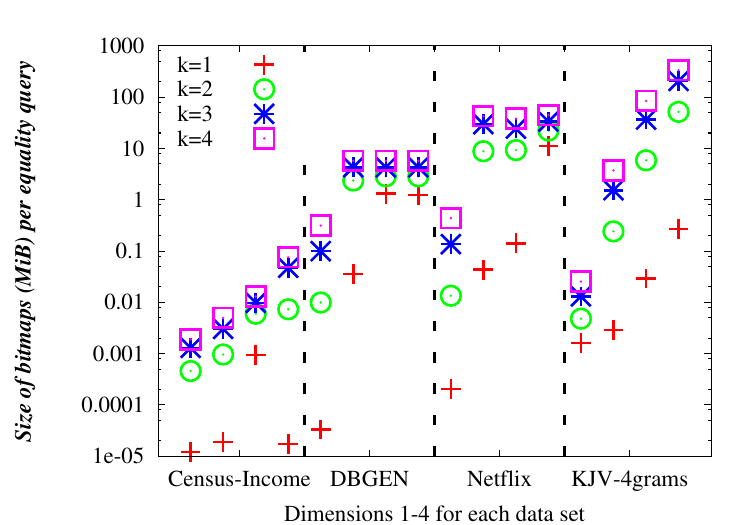}
}
\caption{Bitmap data examined per equality query.
}\label{fig:queries-sizes}
\end{figure}

An exception occurs for the first dimension of Netflix, and it
illustrates the importance of benchmarking with large data sets.  Note that
using $k=1$ is much slower than using $k>1$.  However, these tests were done
using a disk whose access typically\footnote{ This is perhaps pessimistic,
as an operating system may be able to cluster portions of the index for
a given dimension onto a small number of adjacent tracks, thereby reducing
seek times.
} requires at least 18\,ms.
In other words, any query answered substantially faster than 20\,ms was
 answered without retrieving data from the disk
platter (presumably, it came from the operating
system's cache, or perhaps the disk's cache).
For $k>1$, it appears that
the portion of the index for the first attribute  (which is 
$\approx7$ MB\footnote{For $k=2$ we have 981 bitmaps and (see Figure~\ref{fig:queries-sizes}) about 7~kB per bitmap.}) could be cached 
successfully, whereas for $k=1$, the portion of the index was 100\,MB\footnote{
The half-million bitmaps had an average size of about 200 bytes.} and
could not be cached. 

In \S~\ref{sec:multi}, we
predicted that the query time would grow with $k$ as $\approx (2-1/k)n_i^{-1/k}$:
for the large dimensions such as the largest ones for DBGEN (400k)
and Netflix (480k), query times are indeed significantly slower for
 $k=2$ as opposed
to $k=1$. 
However,
our model 
exaggerates
the differences by an order of magnitude. 
\killforgood{Even if we compare CPU time\footnote{\owen{owen: do we need this?}\daniel{I think so, because here we claim that if we were to compare CPU times, we would get some result. A reader could start looking for where we give the numbers. If you want to omit the footnote, then let us omit the discussion about what would happen if we looked at CPU times altogether.} We omit the presentation of CPU times.} rather than wall-clock
time, our model exaggerates by about an order of magnitude.}
The most plausible explanation is that
query times are not 
 proportional to the sizes of the bitmap loaded, but also include
a constant factor.  This may correspond to disk access times.

%

Fig.~\ref{fig:randqueries-times-unsorted} and~\ref{fig:randqueries-times}
also show the equality query times per column before and after sorting the tables.
Sorting improves query times most for larger values of $k$: for Netflix, sorting improved
the query times by  
\begin{itemize}
\item at most 2 for $k=1$, 
\item at most 50  for $k=2$,  
\item and at most 120 for $k=3$.
\end{itemize}
This is consistent with our earlier observation that
indexes with $k>1$ benefit
from sorting even when there are no long runs of identical values (see \S~\ref{sec:sorting-rows}).  (On the first
columns, $k=3$ usually gets the best improvements from sorting.)
The synthetic data set  DBGEN showed no
significant speedup from sorting, beyond its large first column.  Although
Netflix, like DBGEN, has a many-valued column first, it shows a benefit
from sorting even in its third column: in fact, the third column benefits
more from sorting than the second column.  
The largest table, KJV-4grams, benefited most from the sort: while queries
on the last column are up to 10 times faster, the gain on the first two columns 
ranges from 125 times faster ($k=1$) to almost 3\,300 times faster ($k=3$).


We can compare these times with the expected amount of data scanned per query.
This is shown in Fig.~\ref{fig:queries-sizes}, and we observe 
some  
agreement
between most query times and the expected sizes of the bitmaps 
being scanned. 
The most notable exceptions are for $k=1$; in many such cases we must make
an expensive
seek far into a file for a very small compressed bitmap.
Moreover, a small compressed bitmap may, via long runs of 1x11
clean words, represent many row IDs.
To answer the query, we must still produce the set of row IDs. 

\subsection{Effect of the word length}
\label{sec:wordlen-exper}

Our experiments so far use 32-bit EWAH\@. To investigate the effect of
word length, we recompiled our executables as 64-bit binaries and
implemented 16-bit and 64-bit EWAH\@.
The index sizes are reported in Table~\ref{tab:effectofwordlength}---the index size excludes a B-Tree storing maps from attribute values to bitmaps.
We make the following observations:
\begin{itemize}
\item 16-bit indexes can be  10~times larger than 32-bit indexes.
\item 64-bit indexes are nearly twice as large as 32-bit indexes.
\item Sorting benefits 32-bit and 64-bit indexes equally; 16-bit indexes
do not benefit from sorting.
\end{itemize}
Despite the large variations in file sizes,  the difference between index construction times
(omitted) in 32-bit and 64-bit indexes is within 5\%. Hence, index construction
is not bound by disk I/O performance. 

%


\begin{table}
\centering
\caption{Index size (file size in MB) for unary bitmap indexes ($k=1$) under various word lengths. For Census-Income and DBGEN, the 4-d projection is used.
}

\subtable[Unsorted
]{
\begin{tabular}{|c|cccc|cccc|} \cline{2-5}
 \multicolumn{1}{c}{} &  \multicolumn{4}{|c|}{\textbf{index size (MB)}} \\ \hline 
word length	&   Census-Income & DBGEN  & Netflix           & KJV-4grams \\ \hline  
16          &   12.0          & 2.5$\times 10^3$   & 2.6$\times 10^4$  & 2.6$\times 10^4$ \\
32          &    3.8          &	221	   & 2.5$\times 10^3$  & 2.4$\times 10^4$   \\ 
64          &   6.5          & 416    & 4.8$\times 10^3$  & 4.4$\times 10^4$ \\\hline
\end{tabular}
}
\subtable[Lexicographically sorted
]{
\begin{tabular}{|c|cccc|cccc|} \cline{2-5}
 \multicolumn{1}{c}{} &  \multicolumn{4}{|c|}{\textbf{index size (MB)}} \\ \hline 
word length	& Census-Income & DBGEN& Netflix           &KJV-4grams         \\ \hline  
16          & 11.1          & 2.4$\times 10^3$   &  2.5$\times 10^4$  & 1.6$\times 10^4$  \\
32          & 2.9          & 137                 & 1.3$\times 10^3$  & 2.6$\times 10^3$    \\ 
64          & 4.8          &   227               & 2.2$\times 10^3$ & 4.3$\times 10^3$      \\\hline
\end{tabular}
}
\label{tab:effectofwordlength}
\end{table}
%
%
%
%
%
%

\subsection{Range queries}
\label{sec:rangequeries-exper}

Unary bitmap indexes may not be ideally suited for all ranges queries~\cite{1272746}. 
However, range queries are good stress tests: they 
require loading and computing numerous bitmaps. 
Our goal is to survey the effect of sorting and
word length on the aggregation of many bitmaps.

We implemented range queries
using the following simple algorithm:
\begin{enumerate}
\item For each dimension, we compute the logical OR of all
matching bitmaps. We aggregate the bitmaps two at time:
$((B_1 \lor B_2) \lor B_3) \lor B_4) \ldots$ When there are many
bitmaps, Algorithm~\ref{algo:genrunlengthmultiplefaster} or an in-place algorithm might be faster. (See Wu et al.~\cite{1316694,wu2006obi} for a detailed
comparison of
pair-at-a-time versus in-place processing.)
 \item We compute the logical AND of
 all the dimensional bitmaps---resulting from the previous step.
\end{enumerate} 
We implemented a flag to disable the aggregation of the bitmaps
to measure solely the cost of loading the bitmaps in memory.
(Our implementation does not write its temporary results 
to disk.)
We omitted 16-bit EWAH from our tests due to its
poor compression rate.

As a basis for comparison, we also 
implemented range queries using  uncompressed external-memory B-tree~\cite{qdbm} 
indexes over each
column: the index maps values to corresponding row IDs. The computation
is implemented as with the bitmaps, using the STL functions  set\_intersection
and set\_union. We required row IDs to be provided in sorted order.
All query times were at least an order of magnitude
larger than with 32-bit or 64-bit bitmap indexes. 
We failed
to index the columns with  uncompressed B-trees in a 
reasonable time (a week) over the KJV-4grams data set due
to the large file size (21.6\,GB).

We generated a set of uniformly randomly distributed 4-d range queries using
no more than 100~bitmaps per dimension. We used the same set of queries for
all indexes. The results are presented in Table~\ref{tab:effectofwordlengthonrangequeries}.
Our implementation of range queries using uncompressed B-tree indexes is an order of magnitude
slower than the bitmap indexes over Netflix, hence we omit the results.

The disk I/O can be nearly twice as slow with 64-bit indexes and  KJV-4grams. 
However, disk I/O is negligible, accounting for about 1\% of the total time.

The 64-bit indexes are nearly twice as large. 
We expect that 64-bit indexes also generate larger intermediate bitmaps during the computation. Yet, the 64-bit indexes have faster overall performance: 40\% for
DBGEN and 5\% for other cases, except for sorted KJV-4grams where the gain was 18\%. Moreover, the benefits
of 64-bit indexes are present in both sorted and unsorted indexes.

\begin{table}
\caption{Average 4-d  range query processing time over the Netflix data set for unary bitmap indexes ($k=1$) under various word lengths and dimensional B-tree indexes.}

\centering
%
%
\subtable[Average wall-clock query time (s)]{
\begin{tabular}{|c|c|c|} 
\cline{2-3}
 \multicolumn{1}{c}{DBGEN} &  \multicolumn{1}{|c|}{\textbf{unsorted}} & \multicolumn{1}{|c|}{\textbf{lexicographically sorted}} \\  \hline  
32-bit EWAH &    0.382             &   0.378              \\ 
64-bit EWAH &    0.273            &     0.265              \\ 
\hline
 \multicolumn{1}{c}{Netflix} & & \\\hline  
32-bit EWAH &    2.87             &   1.50               \\ 
64-bit EWAH &    2.67            &      1.42              \\ 
\hline
 \multicolumn{1}{c}{KJV-4grams} &  & \\\hline
32-bit EWAH &   44.8             &   5.2               \\ 
64-bit EWAH &    42.4            &      4.4              \\ 
\hline
\end{tabular}
}
\subtable[Average disk I/O time (s)]{
\begin{tabular}{|c|c|c|} 
\cline{2-3}
 \multicolumn{1}{c}{DBGEN} &  \multicolumn{1}{|c|}{\textbf{unsorted}} & \multicolumn{1}{|c|}{\textbf{lexicographically sorted}} \\  \hline  
32-bit EWAH &    0.023             &   0.023              \\ 
64-bit EWAH &    0.027            &     0.026              \\ 
\hline
 \multicolumn{1}{c}{Netflix} & & \\\hline  
32-bit EWAH &   0.11             &        0.078          \\ 
64-bit EWAH &    0.16            &     0.097             \\ 
\hline
 \multicolumn{1}{c}{KJV-4grams} &  & \\\hline
32-bit EWAH &   0.57             &   0.06              \\ 
64-bit EWAH &    1.11            &      0.1              \\ 
\hline
\end{tabular}
}
\label{tab:effectofwordlengthonrangequeries}
\end{table}

\section{Guidelines for k}
\label{sec:Guidelinesfork}

Our experiments indicate that
simple ($k=1$) bitmap encoding is preferable when storage space and
index-creation time are less important than fast equality queries. 
 The storage and index-creation penalties are kept
modest by table sorting and Algorithm~\ref{algo:owengenbitmap}.

Space requirements can be reduced by choosing $k>1$,
although Table~\ref{tab:index-sizes} shows that this approach has risks
(see KJV-4grams). For $k>1$,
we can gain additional index
size reduction 
at the cost of longer index construction 
by using Gray-Frequency rather than Gray-Lex.  \killforgood{The amount of space
gain is difficult to predict. }

If the total number of attribute values is
small relative to
the number of rows, then we should 
first
try the $k=1$ index.  Perhaps the data set resembles
KJV-4grams.  Besides yielding faster queries, 
the $k=1$ index may be smaller.


\section{Conclusion and future work}\label{sec:Conclusion}

We showed that while sorting improves bitmap indexes,
we can improve them even more (30--40\%) if we know the number of distinct values
in each column. For $k$-of-$N$ encodings with $k>1$, even further gains (10--30\%) are
possible using the frequency of each value. 
Regarding future work, 
the accurate mathematical modelling of compressed bitmap indexes
 remains an open problem.
 While we only investigated bitmap indexes,
we can generalize this work in the context of column-oriented databases~\cite{1083658,1142548}
by allowing various types of indexes.



\section*{Acknowledgements}
This work is supported by NSERC grants 155967, 261437 and by  FQRNT 
grant 112381.  
\bibliographystyle{elsarticle-num}
\bibliography{../bib/lemur}

\vspace*{\parsep}
\begin{minipage}{0.2\linewidth}
\includegraphics[width=0.8\columnwidth]{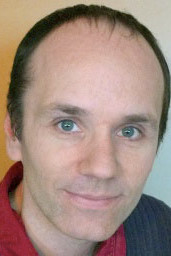}
\end{minipage}
\hfill
\begin{minipage}{0.8\linewidth}
\textbf{Daniel Lemire} received a B.Sc. and a M.Sc. in Mathematics from the University
of Toronto in 1994 and 1995. He received his Ph.D. in Engineering Mathematics from the Ecole Polytechnique and
the Universit\'e de Montr\'eal in 1998. He is now a professor at the Universit\'e du Qu\'ebec \`a Montr\'eal (UQAM)
where he teaches Computer Science. His research interests include data warehousing, OLAP and time series.
\end{minipage}

\vspace*{\parsep}
\begin{minipage}{0.2\linewidth}
\includegraphics[width=0.8\columnwidth]{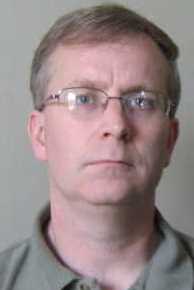}
\end{minipage}
\hfill
\begin{minipage}{0.8\linewidth}
\textbf{Owen Kaser} is Associate Professor in the Department of Computer Science
and Applied Statistics, at the Saint John campus of The University of 
New Brunswick.   He received a Ph.D. in Computer Science in 1993 from
SUNY Stony Brook.
\end{minipage}

\vspace*{\parsep}
\begin{minipage}{0.2\linewidth}
\includegraphics[width=0.8\columnwidth]{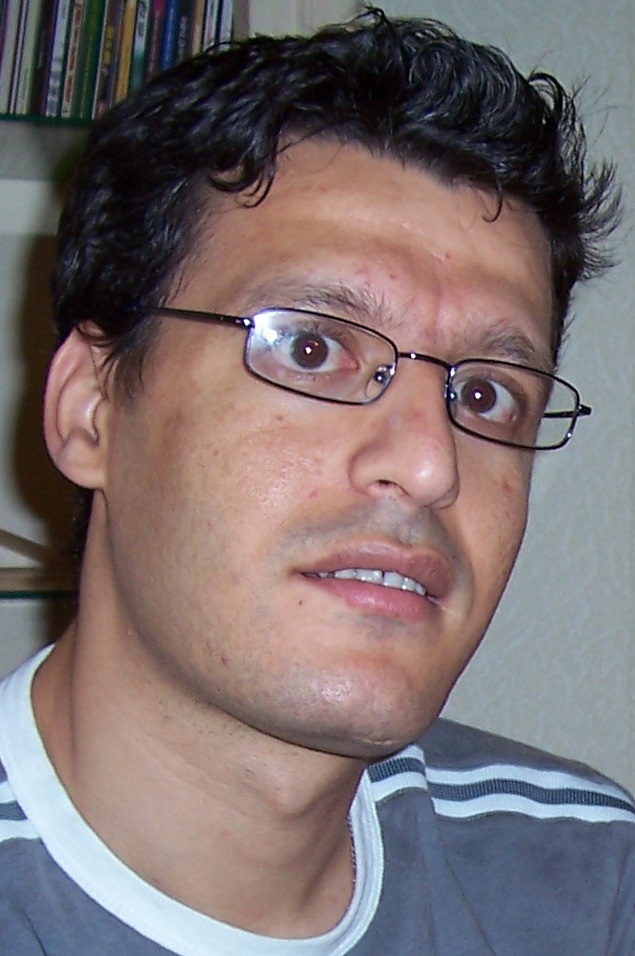}
\end{minipage}
\hfill
\begin{minipage}{0.8\linewidth}
\textbf{Kamel Aouiche} graduated as an engineer from the University Mouloud Mammeri in 1999, he received a B.Sc. in Computer Science from the INSA de Lyon in 2002 and a Ph.D. from the Universit\'e Lumi\`ere Lyon 2 in 2005. He completed a post-doctoral fellowship at Universit\'e du Qu\'ebec \`a Montr\'eal (UQAM) in 2008. Currently, he works as a consultant in industry.
\end{minipage}

\end{document}